\documentclass[]{aa}
\usepackage{graphicx,url,twoopt,siunitx,cases,empheq, hyperref}
\usepackage[version=4]{mhchem}

\hypersetup{colorlinks=true,citecolor=blue} 

\bibpunct{(}{)}{,}{a}{}{,}   
\defcitealias{endalEvolutionRotatingStars1976}{ES}
\defcitealias{coxPrinciplesStellarStructure1968}{CG}
\defcitealias{fabryModelingOvercontactBinaries2022}{Paper I}
\defcitealias{shuStructureContactBinaries1976}{SLA}

\begin{document}

\title{Modeling contact binaries}
\subtitle{II. The effect of energy transfer}
\titlerunning{Modeling contact binaries II. The effect of energy transfer}

\author{
M. Fabry\inst{\ref{ivs}} 
\and P. Marchant\inst{\ref{ivs}}
\and N. Langer\inst{\ref{aifa}, \ref{mpifr}}
\and  H. Sana\inst{\ref{ivs}}
}
\authorrunning{M. Fabry et al.}

\institute{
Institute of Astronomy (IvS), KU Leuven, Celestijnenlaan 200D, B-3001 Leuven, Belgium\\ \email{matthias.fabry@kuleuven.be} \label{ivs} 
\and Argelander-Institut f\"ur Astronomie, Universit\"at Bonn, Auf dem H\"ugel 71, D-53121 Bonn, Germany \label{aifa} 
\and Max-Planck-Institut f\"ur Radioastronomie, Auf dem Hügel 69, D-53121 Bonn \label{mpifr}
}

% \abstract{}{}{}{}{}
% 5 {} tokens are mandatory
\abstract
% context heading (optional)
{
It is common for massive stars to engage in binary interaction.
In close binaries, the components can enter a contact phase, where both stars overflow their respective Roche lobes simultaneously.
While there exist observational constraints on the stellar properties of such systems, the most detailed stellar evolution models that feature a contact phase are not fully reconcilable with those measurements.
% As the contact configuration is inherently three dimensional, simplifying assumptions are made when modeling with one dimensional evolution codes.
% While mass transfer and orbital evolution is taken into account, the tidal deformation on the stellar structure or energy transfer in overcontact layers are ignored.
}
% aims heading (mandatory)
{
We aim to consistently model contact phases of binary stars in a 1D stellar evolution code.
To this end, we develop the methodology to account for energy transfer in the common contact layers.
}
% methods heading (mandatory)
{
We implement an approximative model for energy transfer between the components of a contact binary based on the von Zeipel theorem in the stellar evolution code MESA.
We compare structure and evolution models with and without this transfer and analyze the implications for the observable properties of the contact phase.
}
% results heading (mandatory)
{
Implementing energy transfer helps eliminating baroclinicity in the common envelope between the components of a contact binary, which, if present, would drive strong thermal flows.
We find that accounting for energy transfer in massive contact binaries significantly alters the mass ratio evolution and can extend the lifetime of an unequal mass ratio contact system.
}
% conclusions heading (optional)
{}


\maketitle

\section{Introduction}
Massive stars $(M_{\rm initial} \gtrsim 8 M_\odot)$ are predominantly found in binaries that interact during their lives \citep{sanaBinaryInteractionDominates2012}, which means that the impact of binarity on the evolution of massive stars must be taken into account in stellar structure and evolution models.
This is supported by population synthesis calculations \citep{vanbeverenWROtypeStar1998, deminkIncidenceStellarMergers2014}.
An extreme form of binary interaction happens when the stellar companions enter a contact configuration, where both stars overflow their respective Roche lobes (RLs).
The stars connect at the so-called neck and form a peanut like shape, as determined by the geometry of the Roche potential.
\par
Understanding the evolution of contact systems is of great importance in the study of, among others, stellar mergers, as these are all preluded by an (unstable) contact phase.
Mergers could be the potential source of strong magnetic fields observed in about 7\% of massive stars \citep{ferrarioOriginMagnetismUpper2009, wickramasingheMostMagneticStars2014, fossatiFieldsOBStars2015, schneiderRejuvenationStellarMergers2016, grunhutMiMeSSurveyMagnetism2017b, schneiderStellarMergersOrigin2019, frost2023}.
Furthermore, close binaries are expected to rotate rapidly due to the high Keplerian velocity and quick tidal synchronization.
This might lead to the physical conditions necessary to create long-duration gamma-ray bursts and magnetar-driven superluminous supernovae \citep{yoonEvolutionRapidlyRotating2005, yoonSingleStarProgenitors2006, aguilera-denaRelatedProgenitorModels2018, aguilera-denaPrecollapsePropertiesSuperluminous2020}.
The high rotational velocity could also lead to chemically homogeneous evolution, which is a proposed channel to produce gravitational wave progenitors \citep{marchantNewRouteMerging2016, mandelMergingBinaryBlack2016}.
\par
In the past, the low-mass contact systems, called W Ursae Majoris (W UMa) systems, have been extensively studied.
These systems have an orbital period of less than a day with a total mass on the order of $1 M_\odot$.
Observational campaigns from \citet{eggenContactBinariesII1967} and \citet{binnendijkOrbitalElementsUrsae1970} identified the first of these stars and found that a significant fraction had mass ratios away from unity.
Now, thousands of W UMa systems have been identified thanks to large surveys like the OGLE survey \citep{szymanskiContactBinariesOGLEI2001}.
Observations of massive contact systems however are much rarer, but fortunately, two dozen or so systems have been identified from the VLT Flames Tarantula Survey (VFTS) \citep{evansVLTFLAMESTarantulaSurvey2011}, the MACHO survey \citep{alcockMACHOProjectLMC1997} and OGLE in the Magellanic clouds, as well as studies of galactic targets \citep[e.g.][]{lorenzoMYCamelopardalisVery2014, lorenzoMassiveMultipleSystem2017, yangComprehensiveStudyThree2019}.
\par
Also on the modeling side, most work in the literature is dedicated to understanding the W UMa systems as opposed to massive contact systems.
\citet{kuiperInterpretationLyraeOther1941} initiated theoretical considerations of contact binaries with the argument that due to the stark difference between RL radii and the mass-radius relation of main sequence stars, contact systems with mass ratio away from unity cannot be stable.
The argument of Kuiper transformed into what is now referred to as Kuiper's paradox once a multitude of contact systems were observed with unequal mass components.
However, it should be clear immediately that this paradox is only apparent as the mass-radius relation of single main sequence stars need not apply for stars in a contact configuration.
\par

\citet{lucyStructureContactBinaries1968} first considered energy transfer (ET) in common convective envelopes of stellar components, and computed the first approximated structure models for W UMa stars.
Later, \citet{lucyUrsaeMajorisSystems1976, flanneryCyclicThermalInstability1976, hazlehurstDissipationFactorContact1985} and \citet{kahlerStructureEquationsContact1989} calculated models that are out of thermal equilibrium and show cyclic behavior.
\citet{shuStructureContactBinaries1976, shuStructureContactBinaries1979} and \citet{lubowStructureContactBinaries1977, lubowStructureContactBinaries1979} (collectively \citetalias{shuStructureContactBinaries1976}) constructed models of contact binaries where they dropped the requirement of a continuous structure at the layer coinciding with the equipotential surface of the first Lagrangian point ($L_1$).
These models resolve Kuiper's paradox regardless of the thermal structure of the envelopes, be they radiative or convective.
Despite much criticism \citep[see e.g.][]{hazlehurstEquilibriumContactBinary1993, kahlerStructureEquationsContact1989}, this is the simplest model of ET in radiative envelopes present in the literature.
\par

Accurately modeling massive, long-lived contact systems has been attempted in the past.
\citet{marchantNewRouteMerging2016} computed detailed evolution model grids of massive binaries with initial periods down to $\SI{0.5}{d}$, which includes the regime of contact binaries at the zero age main sequence.
Those models however did not include energy transfer between contact components, as the models concerned binaries of mass ratio close to unity $M_2/M_1 = \numrange{0.8}{1}$ and the effect was thought to be small.
\par
\citet{senDetailedModelsInteracting2022} used the models of \citet{marchantImpactTidesMass2016} to study the semi-detached Algol systems, although these models also included contact phases.
\citet{menonDetailedEvolutionaryModels2021} extended these grids by computing models with initial mass ratios down to $M_2/M_1 = 0.6$ with the study of massive, long-lived contact binaries in mind.
They found, also without including energy transfer in contact phases, a strong correlation between observed mass ratio and period in contact systems, broadly in agreement with observation.
However, the mass ratio distribution they derive is heavily skewed toward values close to unity, which is not supported observationally.
They suggest that including energy transfer in contact phases of unequal mass components could alleviate this discrepancy.
\par

In this work, we apply the ET model of \citetalias{shuStructureContactBinaries1976} in common stellar layers to modern stellar structure models, to further advance our evolutionary modeling of massive contact binaries.
In Sect. \ref{sec:ettheory}, we describe and discuss the theory of ET.
Section \ref{sec:meth} describes the physical setup of the stellar evolution code, along with our ET implementation.
In Sect. \ref{sec:models}, we compare models computed with and without ET in contact layers and discuss the difference in the observable properties of the models.
Lastly, in Sect. \ref{sec:conc}, we provide our concluding remarks.

\section{Theory of energy transfer}\label{sec:ettheory}

\subsection{Simple considerations}
The theoretical modeling of contact stars started with \citet{kuiperInterpretationLyraeOther1941}, who stated that stable contact systems of uniform composition with unequal masses cannot exist as a result of differing mass-radius relationships. 
For the galactic, zero age main sequence (ZAMS) models of \citet{brottRotatingMassiveMainsequence2011}, we derive the mass radius relation of single stars to be approximately
\begin{equation}\label{eq:massradius}
    \left.\frac{R_2}{R_1}\right|_{\rm ZAMS} = \left(\frac{M_2}{M_1}\right)^{0.57} = q^{0.57},
\end{equation}
where we defined $q\equiv M_2/M_1$.
However, the condition of contact in a binary following the Roche geometry constrains the surface of the stars to same equipotential, which leads to
\begin{equation}\label{eq:contactapprox}
    \left.\frac{R_2}{R_1}\right|_{\rm Roche} \approx \left(\frac{M_2}{M_1}\right)^{0.46} = q^{0.46},
\end{equation}
for stars not overflowing their RL too much.
Clearly, both these equations \eqref{eq:massradius}-\eqref{eq:contactapprox} cannot be satisfied simultaneously unless $M_1 = M_2$.
However, while the contact condition needs to be satisfied from dynamical arguments, the stellar structure need not, a priori, follow a single star model.
Even though the dense stellar core will be largely unperturbed due to the companion, the outer layers of contact components are highly distorted, causing different total radii with respect to spherical models, as seen explicitly in \citet{fabryModelingOvercontactBinaries2022} (henceforth \citetalias{fabryModelingOvercontactBinaries2022}).
This result did not yet take energy transfer into account, and we expect further changes under the consideration that a hotter gas under similar pressure takes up more volume.
Therefore, the ZAMS mass-radius relation of Eq. \eqref{eq:massradius} is not expected to be satisfied under general contact conditions, and so Kuiper's paradox can be resolved by providing alternative stellar models that have a mass-radius relation closer to the contact condition of Eq. \eqref{eq:contactapprox}.
\par
Following the Roche lobe geometry, \citet{lucyStructureContactBinaries1968} finds the ratio of the surface areas $S_2/S_1$ of two stars in contact to be proportional to $(M_2/M_1)^{\beta}$, with $\beta=0.96$. 
Using the approximation $\beta \simeq 1$, combined with von Zeipel's theorem of gravity darkening $T_{\rm eff}^4 \propto g$ \citet{vonzeipelRadiativeEquilibriumRotating1924}, and the Stefan-Boltzmann law $L \propto S T_{\rm eff}^4$, results in the simple expectation that in contact binaries, the luminosity ratio follows the mass ratio \citep{lucyStructureContactBinaries1968, tassoulStellarRotation2000}:
\begin{equation}\label{eq:contactML}
    \frac{L_2}{L_1} \simeq \frac{M_2}{M_1} = q.
\end{equation}
Single main sequence stars, on the other hand, follow the well known mass-luminosity relation 
\begin{equation}\label{eq:singleML}
    \frac{L_{\rm s, 2}}{L_{\rm s, 1}} \simeq \left(\frac{M_2}{M_1}\right)^{\alpha} = q^\alpha,
\end{equation} 
with $\alpha\simeq\numrange{2}{3}$ for the upper main sequence \citep{kohlerEvolutionRotatingVery2015, grafenerEddingtonFactorKey2011}, and we use the symbol $\simeq$ to denote these relations are approximations to simple power laws.
We see that this leads to a difference between the luminosity of a single star $L_{s,1}$ and the luminosity $L_1$ of a star of the same mass in a contact binary of
\begin{equation}\label{eq:dL1}
    \Delta L_1 = L_1 - L_{1,s} = -f L_1,
\end{equation}
and for the companion
\begin{equation}\label{eq:dL2}
    \Delta L_2 = L_2 - L_{2,s} = f L_1 \simeq \frac{f}{q} L_2,
\end{equation}
since $L_2\simeq q L_1$ and we require $\Delta L_1 + \Delta L_2 = 0$ to conserve energy.
This defines $f$ as
\begin{equation}\label{eq:dLf}
f\simeq\frac{q-q^{\alpha}}{1+q^{\alpha}}.
\end{equation}
Therefore, the two stars in a contact binary can fulfill the single star mass-luminosity relation of Eq. \eqref{eq:singleML} in their cores and the contact binary mass-luminosity condition of Eq. \eqref{eq:contactML} at their surfaces if the amount of energy per time given by Eqs. \eqref{eq:dL1} and \eqref{eq:dL2} are transferred from the more massive to the less massive star in their common envelope.

\subsection{Models of energy transfer}
The general solution to Kuiper's paradox is to consider detailed stellar models with the inclusion of energy transfer (ET) between the binary components.
There exist several models of ET in the literature.
\par
\citet{lucyStructureContactBinaries1968} and \citet{biermannModelsContactBinaries1972} provided a first solution by adjusting the adiabatic constants of convective envelopes in contact components.
However, this is an unsatisfactory solution, since this setup requires the stars to be burning hydrogen through different nuclear chains or cycles in the case of \citet{lucyStructureContactBinaries1968}, or that the models exhibit inaccurate light curves in \citet{biermannModelsContactBinaries1972}.
Other models like those of \citet{lucyUrsaeMajorisSystems1976} or \citet{ flanneryCyclicThermalInstability1976} relaxed the requirement of thermal equilibrium and constructed models of W UMa stars that exhibited thermal cycles. 
\citet{kahlerStructureEquationsContact1989} presented a detailed model that required turbulent motions in the common envelope to explain early type (radiative) W UMa binaries.
\par
Meanwhile, \citetalias{shuStructureContactBinaries1976} presented the contact discontinuity model of contact binaries, by relaxing the requirement of continuous structural quantities across the RL.
This is the only model that treats the common envelope as a single volume of the binary structure, at the price of hiding a heat engine in a very thin region around the RL.
One peculiar feature is that this model necessitated a temperature inversion at the RL layer in one of the components as otherwise they would not be able to construct thermally stable contact models of uniform composition (in order to model binaries at zero age). 
This feature has received criticism in that the proposed heat engine violates the second law of thermodynamics and cannot be stable over thermal timescales \citep[see][and references therein]{hazlehurstEquilibriumContactBinary1993, kahlerStructureEquationsContact1989}.
\par
Later, \citet{kahlerStructureContactBinaries2004} concludes from all collected theoretical arguments that internal circulation currents must exist in the less luminous component that reduces the radiative temperature gradient since the luminosity carried by radiation is reduced by the circulation luminosity.
\par
Given the complexity of the theoretical problem of the structure of contact binaries, especially with radiative envelopes, it is beyond the scope of this work to further develop the analytic theory.
Instead, we apply an ET model in modern stellar structure calculations.
With shellularity as our base assumption of the stellar structure (see Sect. \ref{ssec:roche} for the precise definition), the model of \citetalias{shuStructureContactBinaries1976} is a natural choice, though we recognize that this comes with the apparent thermodynamical problems stated above.
However, we believe we avoid the most fundamental one, as we do not explicitly require a temperature inversion in our models.
We only use the model of \citetalias{shuStructureContactBinaries1976} to compute the amount of energy transferred (see Sect. \ref{ssec:ctctdis}).

\subsection{Energy transfer in the Roche geometry}\label{ssec:ctctdis}
The work of \citetalias{shuStructureContactBinaries1976} provides a general model of ET, by introducing the notion of an energy flow at the base of the common envelope.
If the contact layers are shellular, and satisfy von Zeipel's gravity darkening, conservation of energy at the RL implies:
\begin{subequations}\label{eq:newL}
\begin{align}
    L_1' &= (L_1+L_2) \frac{S_1\langle g\rangle_1}{S_1\langle g\rangle_1 + S_2\langle g\rangle_2},\\
    L_2' &= (L_1+L_2) \frac{S_2\langle g\rangle_2}{S_1\langle g\rangle_1 + S_2\langle g\rangle_2}.
\end{align}
\end{subequations}
Here, $S$ is the surface area of the RL, $\langle g\rangle$ the surface averaged effective gravity at the RL, and the primed quantities specify the state just above the ET layer, while unprimed those just below.
This equation specifies that the fraction of the total luminosity that each component radiates is proportional to $S\langle g\rangle$.
The transferred luminosity then equals
\begin{equation}\label{eq:etransfer}
    L_{\rm trans} = L_1 - L_1' = \frac{L_1 S_2\langle g\rangle_2 - L_2 S_1\langle g\rangle_1}{S_1\langle g\rangle_1 + S_2\langle g\rangle_2}.
\end{equation}
Comparing Eq. \eqref{eq:newL} against Eqs. \eqref{eq:dL1}-\eqref{eq:dL2}, we find for the fraction $f$:
\begin{equation}
    f = \frac{L_1S_2\langle g\rangle_2-L_2S_1\langle g\rangle_1}{(L_1+L_2)S_1\langle g\rangle_1},
\end{equation}
which is consistent with Eq. \eqref{eq:dLf} as $\frac{S_2\langle g\rangle}{S_1\langle g\rangle} \simeq q$ and $\frac{L_2}{L_1}\simeq q^\alpha$.
\par

\section{Methods}\label{sec:meth}
To investigate the effect of ET on the evolution of contact binaries, we compute binary evolution models using the stellar evolution code MESA \citep{paxtonModulesExperimentsStellar2011, paxtonModulesExperimentsStellar2013, paxtonModulesExperimentsStellar2015, paxtonModulesExperimentsStellar2018, paxtonModulesExperimentsStellar2019, jermynModulesExperimentsStellar2023}, version r22.11.1.
% , with the \texttt{mesasdk-x86\_64-macos-22.6.2} SDK.
These models are the first massive binary evolution models that include ET in contact layers.
We follow the evolution of binaries from the ZAMS until the least massive component overflows the second Lagrangian point.

\subsection{Physical assumptions in MESA}
The microphysical setup of our stellar evolution models remains mostly the same as in \citetalias{fabryModelingOvercontactBinaries2022}, with some additions from the newer MESA version.
In summary, the nuclear net contains the eight isotopes of $\ce{^{1}H, ^{3}He, ^{4}He, ^{12}C, ^{14}N, ^{16}O, ^{20}Ne}$ and $\ce{^{24}Mg}$, sufficient for the main sequence and nuclear burning rates are taken from the JINA \citep{cyburtJINAREACLIBDatabase2010} and NACRE libraries, with weak interaction rates from \citet{fullerStellarWeakInteraction1985, odaRateTablesWeak1994} and \citet{langankeShellmodelCalculationsStellar2000}.
Plasma screening is included following \citet{chugunovCoulombTunnelingFusion2007} and thermal neutrino losses are computed in \citet{itohNeutrinoEnergyLoss1996}.
For the equation of state, a blend is used between different tables by \citet{saumonEquationStateLowMass1995, timmesAccuracyConsistencySpeed2000, rogersUpdatedExpandedOPAL2002, potekhinThermodynamicFunctionsDense2010} and \citet{jermynSkyeDifferentiableEquation2021}, specified in \citet{jermynModulesExperimentsStellar2023}.
Radiative opacities are also blended from CO-enhanced tables of OPAL opacities \citep{iglesiasRadiativeOpacitiesCarbon1993, iglesiasUpdatedOpalOpacities1996} and \citet{fergusonLowTemperatureOpacities2005} for lower temperatures.
At high temperatures, Compton scattering opacity is from \citet{poutanenRosselandFluxMean2017}, and electron conduction opacities are taken from \citet{cassisiUpdatedElectronconductionOpacities2007} and \citet{blouinNewConductiveOpacities2020}.
In all simulations in this work, we set the metallicity of stars to the solar value $Z_\odot$, where $Z_\odot = 0.0142$, with metal fractions as determined from \citet{asplundChemicalCompositionSun2009}.
\par
Mass loss through winds is accounted for by following the prescription of \citet{brottRotatingMassiveMainsequence2011}.
If the surface hydrogen fraction is $X > 0.7$, the mass loss rate is taken either from \citet{vinkMasslossPredictionsStars2001} for temperatures above the iron bi-stability jump \citep[calibrated also by ][]{vinkMasslossPredictionsStars2001} or the maximum of the rates \citet{vinkMasslossPredictionsStars2001} and \citet{nieuwenhuijzenAtmosphericAccelerationsStability1995} below this temperature.
For $X < 0.4$, the wind prescription of \citet{hamannSpectralAnalysesGalactic1995} is used, although decreased by a factor of ten.
When $0.4 < X < 0.7$, the wind is linearly interpolated between the above results.
We allow part of the wind launched by a star to be accreted by its companion using the Bondi-Hoyle mechanism \citep{bondiMechanismAccretionStars1944} as implemented by \citet{hurleyEvolutionBinaryStars2002}.
\par
For mixing, we use the Ledoux criterion to determine convectively mixed regions, where the mixing length theory of \citet{bohm-vitenseUberWasserstoffkonvektionszoneSternen1958}, in the version described by \citet{coxPrinciplesStellarStructure1968}, is applied with a mixing length parameter of $\alpha = 2$.
The convective core is allowed to overshoot its boundary.
Following the calibration of \citet{brottRotatingMassiveMainsequence2011}, we use a step overshoot where the diffusion coefficient 0.01 pressure scale heights into the convective layer is kept constant out to 0.335 pressure scale heights beyond the boundary. 
Semiconvection is included following the model of \citet{langerSemiconvectiveDiffusionEnergy1983} with a high efficiency of $\alpha_{\rm sc} = 100$ as calibrated by \citet{schootemeijerConstrainingMixingMassive2019}.
We include thermohaline mixing as developed by \citet{kippenhahnTimeScaleThermohaline1980} with an efficiency parameter $\alpha_{\rm th} = 1$.
All other mixing processes, in particular rotational mixing, are ignored in order to isolate as best as possible the effect of energy transfer.
\par
At all times, the rotation of the stars is synchronized to the orbital period, as part of the requirement to use the Roche potential as the deformation geometry (Sect. \ref{ssec:roche}).
Additionally, we artificially diffuse the total angular momentum throughout the interior of the star to enforce solid body rotation.
The treatment of mass transfer (MT) is explained in Sect. \ref{ssec:mt}, while the implementation of energy transfer (ET) is shown in Sect. \ref{ssec:et}.

\subsection{Shellularity and Roche Lobe geometry}\label{ssec:roche}
Since we deal with highly tidally deformed stars, we use the modifications to the stellar structure equations from \citetalias{fabryModelingOvercontactBinaries2022} to incorporate the RL geometry into a one-dimensional (1D) stellar evolution code.
The stars are therefore modeled as hydrostatic structures living in the Roche potential $\Psi$ of a fully synchronized binary \citepalias[see Eq. 19 of][]{fabryModelingOvercontactBinaries2022}.
Furthermore, we assume stellar layers to be shellular.
Shellularity is reached when all intensive quantities, in particular the temperature, pressure and mass density are constant along a stellar layer, and that such a layer coincides with a unique equipotential surface.
\par
It should be emphasized that, in context of 1D models, full shellularity of layers of a (single) star is an assumption, and not a self-consistently modeled feature as of course one needs a 3D structure to explicitly test shellularity of a stellar layer.
However, overflowing layers of components in a contact binary can be tested to be shellular with each other if the temperature, density, etc. at the 1D cells are equal for equal values of Roche potential. 
Given the method we used to split the common envelope of the contact binary (Sect. 2.4 of \citetalias{fabryModelingOvercontactBinaries2022}), the computation of the tidal deformation corrections in \citetalias{fabryModelingOvercontactBinaries2022} based on equipotential surfaces and the usage of the corresponding outer boundary condition (Sect. 4 of \citetalias{fabryModelingOvercontactBinaries2022}), the final ingredient of ET in the common envelope developed here will ensure shellularity of one component with the other, again, defined as having similar cell values of pressure, temperature, etc. for similar values of potential.
\par
In contact, stellar layers are shared between the components and a choice must be made as to what part of the envelope belongs to each component.
To this end we constructed what we call a splitting surface emanating from the Lagrangian point ${\rm L}_1$, by following the gradient $\nabla\Psi$ outward in all directions (see \citetalias{fabryModelingOvercontactBinaries2022}, Sect. 2.4).
With this setup, the volume of both components is uniquely defined, and in \citetalias{fabryModelingOvercontactBinaries2022} we computed integrals in the Roche potential necessary to represent distorted shells in 1D.
The volume equivalent radius then acts as the new independent variable of a stellar shell, and is defined as:
\begin{equation}
    V_\Psi \equiv \frac{4\pi}{3}r_\Psi^3.
\end{equation}
Given a certain splitting strategy, and the requirement of the stellar surface being on a common equipotential, a relation of the form
\begin{equation}\label{eq:contact}
    r_{\Psi, 2} = F(q; r_{\Psi, 1}),
\end{equation}
with $q = M_2 / M_1$, constrains the radii of the overflowing components 1 and 2. 
This is a statement of the contact condition, and is at the basis of Kuiper's paradox (Sect. \ref{sec:ettheory}).
For vertical splitting surfaces through ${\rm L}_1$, \citet{marchantNewRouteMerging2016} found the fit:
\begin{equation}
    \frac{r_{\Psi, 2} - r_{\rm RL, 2}}{r_{\Psi, 2}} \approx q^{-0.52}  \frac{r_{\Psi, 1} - r_{\rm RL, 1}}{r_{\Psi, 1}}, 
\end{equation}
where we have denoted $r_{\rm RL} = r_{\Psi_{{\rm L}_1}}$ the volume equivalent RL radius of a component.
With our setup of the splitting surface however, this approximation is no longer accurate for significantly overflowing shells of mass ratios away from unity.
Therefore, in this work, we evaluate the function $F$ of Eq. \eqref{eq:contact} by interpolating the results of the Roche integrations obtained in \citetalias{fabryModelingOvercontactBinaries2022}.
\par
Retaining the form of the fit of \citet{marchantRoleMassTransfer2021}, for the radius of a component overflowing to its outer Lagrangian point ${\rm L}_{\rm out}$, which is ${\rm L}_2$ for the less massive and ${\rm L}_3$ for the more massive component, we construct a new fit from the integration results of \citetalias{fabryModelingOvercontactBinaries2022}, which is:
\begin{subequations}\label{eq:l2fit}
    \begin{align}
        \frac{r_{L_{\rm out}} - r_{\rm RL}}{r_{\rm RL}} &= \frac{3.3752}{1+\left(\frac{\ln q + 1.0105}{\sigma}\right)^2}\cdot \frac{1}{9.0087+q^{-0.4022}},\\
        \sigma &= \frac{62.9237}{15.9839 + q^{0.2240}}.
    \end{align}
\end{subequations}

This fit has an error smaller than 0.1\% in the range $-7 \leq \log q \leq 7$, compared to 1\% if we used the fit of \citet{marchantRoleMassTransfer2021}.
We use this fit then to determine when the least massive component overflows to ${\rm L}_2$, after which we stop the evolutionary simulation.
\par
% In Appendix \ref{app:convergence}, we test for convergence of our models given 
\subsection{Mass transfer}\label{ssec:mt}
The components of a contact binary are, per definition, in mechanical contact with each other. Thus, stellar material can be exchanged throughout the evolution of such a system.
However, MT cannot occur at an arbitrary rate, since the surface of both components are constrained by the contact condition of Eq. \eqref{eq:contact}.
If this were not satisfied, pressure gradients would arise across the surface of the contact binary, and we expect this would induce strong, horizontal\footnote{Horizontal in the context of this work means perpendicular to the local effective gravity $\vec{g} = \nabla\Psi$. Conversely, vertical means parallel to $\nabla\Psi$.} flows equilibrating the surface.
Thus, in our simulations, we implicitly adjust the MT rate either to keep the donor just below its RL radius in a semidetached configuration, or so the component radii satisfy the contact condition of Eq. \eqref{eq:contact} in a contact configuration \citep[see also][]{marchantNewRouteMerging2016}.
\par

\subsection{Energy transfer}\label{ssec:et}
Not only are the contact layers of both stellar components in mechanical contact, facilitating MT, there is also thermal contact between these layers, allowing for ET.
As elaborated in Sect. \ref{ssec:ctctdis}, we model ET in contact binaries as the transfer of luminosity of one component to the other (Eq. \eqref{eq:etransfer}), at the location of the RL.
It should be mentioned that no further corrections to the ET need to be computed since the splitting surface we use, computed in \citetalias{fabryModelingOvercontactBinaries2022}, is parallel to the local effective gravity and thus the radiative flux.
Therefore, since the regular energy transport by radiation (or by convection) in the interior of either component is vertical, no vertical energy flux crosses over toward the other component.
Our splitting strategy thus nicely decouples horizontal and vertical energy flows and in this approximation, the ET computed here is not affected by the vertical energy transport present in the components.
\par
The stability and performance of numerical simulations are severely impacted by introducing sharp discontinuities at the RL radius, in this case that of luminosity.
To remedy these issues, we use the following approach.
\par
First, for numerical stability over evolutionary steps, we smooth the amount of ET by computing
\begin{equation}
    \tilde{L}_{\rm trans}^{n} = p L_{\rm trans}^{n} + (1-p) \tilde{L}_{\rm trans}^{n-1},
\end{equation}
where $\tilde{L}_{\rm trans}^{n}$ is the smoothed transferred luminosity at evolutionary step $n$, $L_{\rm trans}$ is the luminosity to be transferred as per the Roche geometry (Eq. \eqref{eq:etransfer}), and $p$ is the smoothing factor.
In all simulations including ET, we use a moderate smoothing of $p=0.5$.
\par
Then, we implement the luminosity transfer as a constant extra source of specific heat $\varepsilon_{\rm RL, 1, 2}$ in both stars, occurring in the cells with a radius within 1.00 and 1.01 times its respective RL radius, a choice so as to reflect the estimate of \citet{shuStructureContactBinaries1979} (see Eq. \eqref{eq:etthickness} below and also Appendix \ref{app:et} for further details).
The magnitude of $\varepsilon_{\rm extra, 1, 2}$ is determined from the required luminosity:
\begin{subequations}\label{eq:heats}
\begin{align}
    \varepsilon_{\rm RL, 1} \Delta m_1 &= \tilde{L}_{\rm trans},\\
    \varepsilon_{\rm RL, 2} \Delta m_2 &= -\tilde{L}_{\rm trans},
\end{align}
\end{subequations}
where $\Delta m_{1, 2}$ is the mass of the cells we put the extra heat in.
\par
Since MT is modeled as removing stellar material from the top of the donor and putting it on top of the accretor, during MT the common layers are out of thermal equilibrium due to expansion or compression.
To correct for this effect and regain shellularity, especially at the surface, we transport additional heat in the common envelope according to Eq. \eqref{eq:etransfer}, where now we evaluate $S\langle g\rangle $ at the surface layer of both components instead of the RL.
The constant heat source $\varepsilon_{\rm surf, 1, 2}$ is then computed similarly to Eq. \eqref{eq:heats}, only now $\Delta m$ encompasses all shells from 1.01 times the RL up to the surface.
\par
As a final approximation, we linearly scale the necessary ET for shallow contact systems, defined as when $0 \leq \frac{R-R_{\rm RL}}{R_{\rm RL}} \leq 0.01$ for either star:
\begin{equation}
    \tilde{L}_{\rm trans, shallow} = \tilde{L}_{\rm trans} \frac{\min\left(\frac{R_1 - R_{\rm RL, 1}}{R_{\rm RL, 1}}, \frac{R_2 - R_{\rm RL, 2}}{R_{\rm RL, 2}}\right)}{0.01}.
\end{equation}
\par

We recognize that using our ET prescription in fast MT phases is the weakest element of our simulations.
When the MT timescale is shorter than the thermal timescale, the star falls out of thermal equilibrium, which is characterized by strong (vertical) gradients in the luminosity profile.
During such phases then, the energy budget of the star is redistributed vertically in order to regain as best as possible thermal equilibrium.
Adding to this process the horizontal energy transport in the layers near the RL could induce complex interactions between both energy flows.
Even though we mentioned above the horizontal and vertical energy flows to be decoupled, this is an idealization and the approximation may break down in quickly evolving phases.
Furthermore, while our computation of how much energy is transferred depends only on the geometry and the theorem of von Zeipel, which does not require thermal equilibrium of the layers, hydrodynamical effects break the validity of von Zeipel's theorem \citep{vonzeipelRadiativeEquilibriumRotating1924}, which can be significant near the Lagrangian point ${\rm L}_1$.
\par

\section{Stellar models}\label{sec:models}

\subsection{Detailed example of ET evolution}\label{ssec:detailed}
Here we detail the evolution of a $25 M_\odot$ primary and a $20 M_\odot$ secondary (so $q_{\rm ini}=0.8$) in an initial orbit of 1.4 days.
This initial configuration is chosen so that, at the onset of the long lasting contact phase, this model mimics the observed galactic contact binary V382 Cyg studied in \citet{abdul-masihConstrainingOvercontactPhase2021}.
We simulate both cases where ET, as implemented in Sect. \ref{ssec:et}, is included and excluded.
\par
\begin{figure*}
    \centering
    \includegraphics[width=1.5\columnwidth]{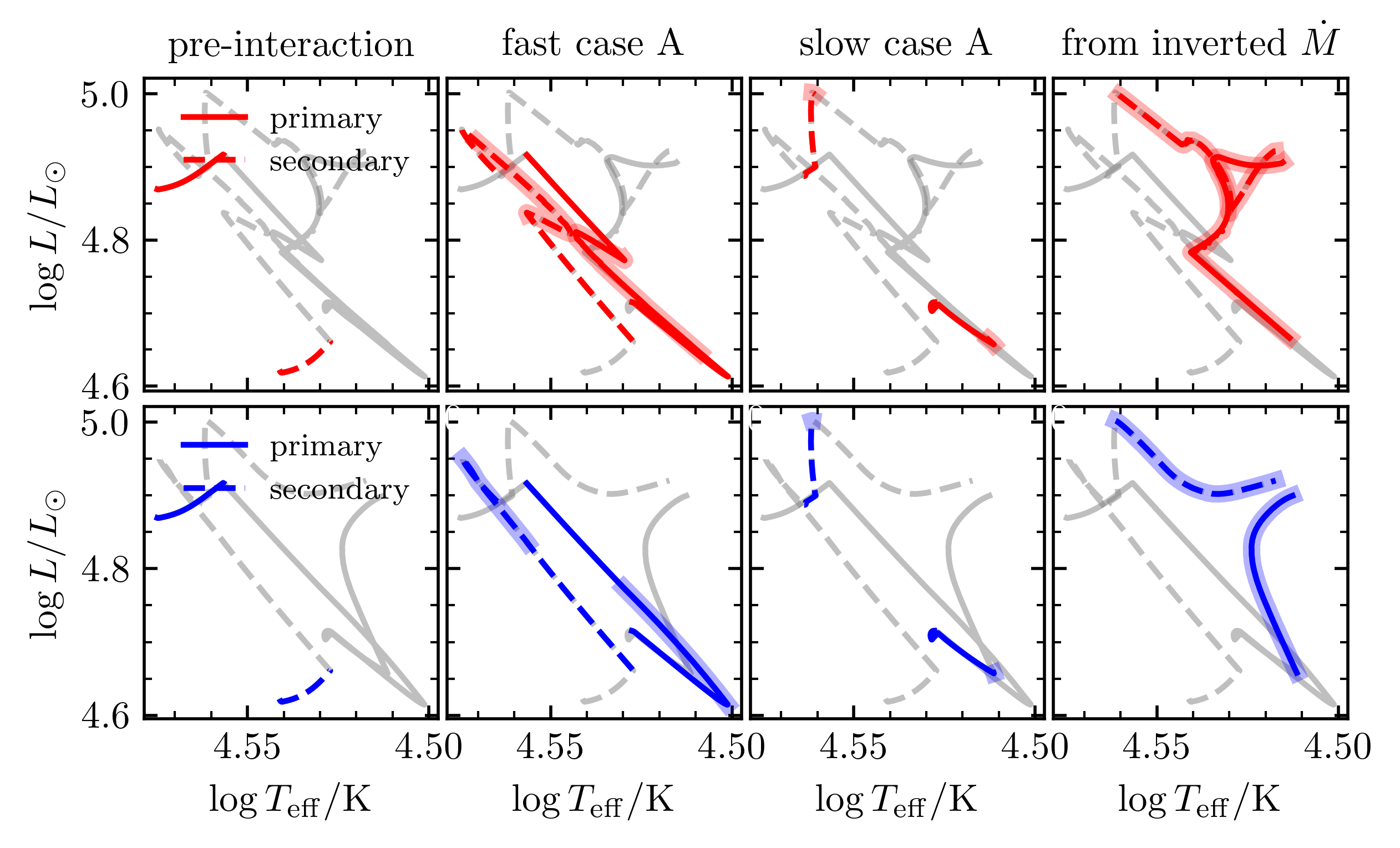}
    \caption{HRD of a 25 + 20 $M_\odot$ binary with an initial period of $1.4\si{d}$. Red lines indicate a simulation with the inclusion of ET, while blue lines ignore ET. Moving left to right, the columns highlight four subsequent phases in the evolution (see Sect. \ref{ssec:detailed} for details). Contact phases are highlighted with thick outlines.}
    \label{fig:hrd}
\end{figure*}
\begin{figure}
    \centering
    \includegraphics{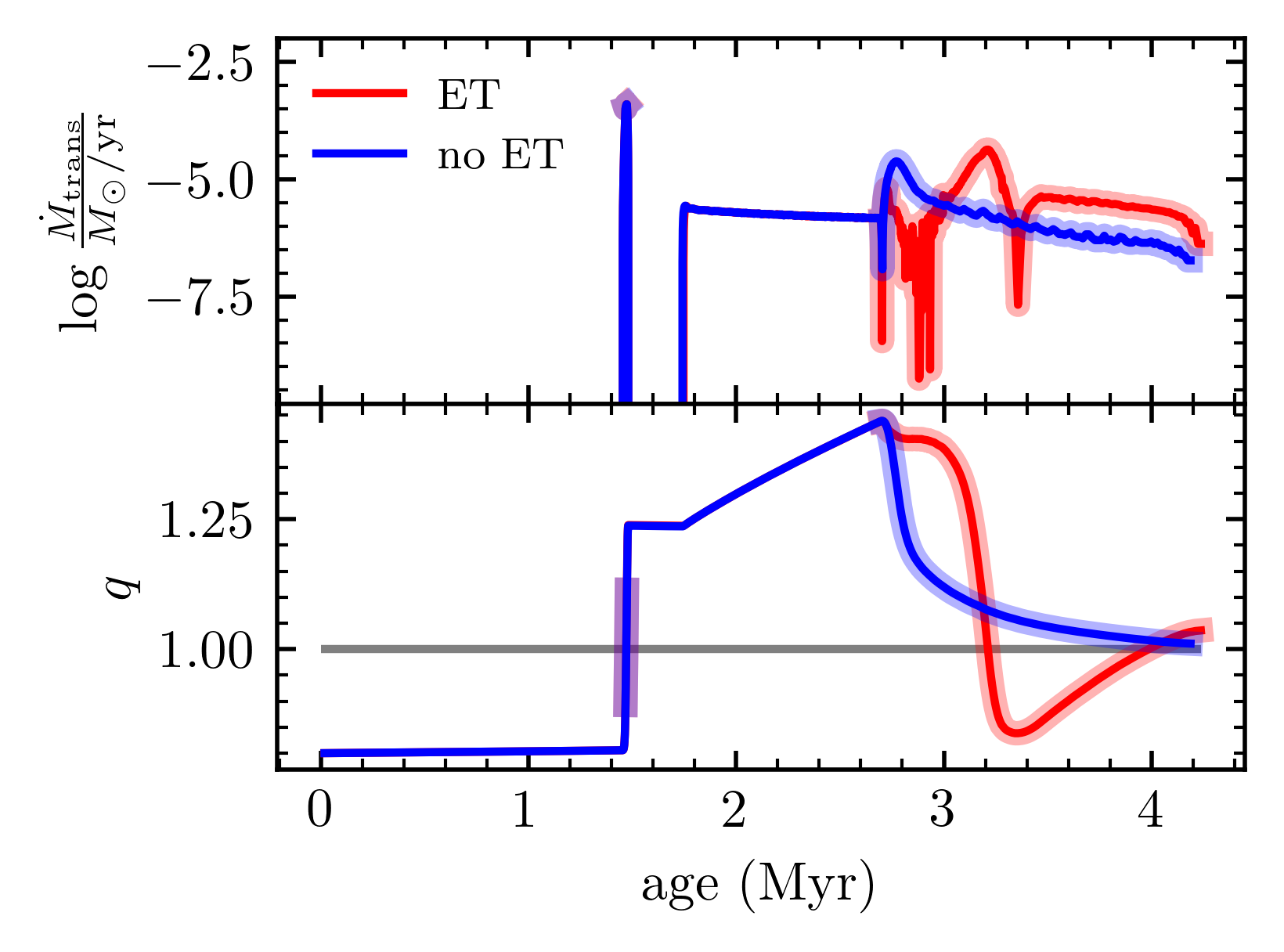}
    \caption{MT rate (top panel) and mass ratio (bottom panel) evolution of the 25 + 20 $M_\odot$ binary studied in Sect. \ref{ssec:detailed}. Contact phases are highlighted with thick outlines.}
    \label{fig:mdot}
\end{figure}

In Fig. \ref{fig:hrd}, we show the Hertzsprung-Russel diagram (HRD) broken up into four distinct phases of the evolution, while in Fig. \ref{fig:mdot}, we show the MT rate and mass ratio evolution.
First, we have the pre-interaction phase (leftmost column) followed by a short lived, fast case A MT phase from the primary to the secondary when the donor experiences RL overflow (second column of Fig. \ref{fig:hrd}).
During this MT phase, the components briefly enter into contact, as the accretor falls highly out of thermal equilibrium and swells beyond ${\rm L}_1$.
After thermal relaxation of the accretor, the system detaches, after which slow case A MT is initiated (third column of Fig. \ref{fig:hrd}), still from the primary to the secondary, although the mass ratio has inverted by this point.
This phase of nuclear timescale MT occurs in a semi-detached configuration and is referred to as an Algol phase.
Finally, the evolution of the now more massive secondary catches up to that of the stripped primary, contact is engaged again and MT inverts (rightmost column of Fig. \ref{fig:hrd}), moving material from the secondary back to the primary (however see below).
This contact configuration evolves on the nuclear timescale and ends when ${\rm L}_2$ overflow occurs.

\par

Qualitatively, the evolution of this system is similar with and without inclusion of ET.
Up to and including the slow case A MT phase, the evolution of this system is nearly identical. 
One exception is the different HRD tracks during the fast MT case.
In the second column of Fig. \ref{fig:hrd}, the red tracks exhibit a `hook', while the blue tracks do not.
This is a direct consequence of ET in the short contact phase changing the outer structure of the components.
The secondary donates energy to the primary and cools a bit, jumping to the right of the HRD (and vice-versa for the primary).
The point at which the primary and secondary tracks touch (at roughly $\log L/L_\odot=4.8$ and $\log T_{\rm eff}/K=4.545$) corresponds to the situation where $M_1 = M_2$.
This is consistent with the theory since we expect that for $M_1=M_2$ we have similar luminosities per Eq. \eqref{eq:contactML} and also similar effective temperatures following the law of Stefan-Boltzmann.
When contact is disengaged, the secondary (at this point more massive than the primary) jumps back to the left as the energy sink to the primary disappears.
Given the timescale of this MT phase however, it is unlikely to observe contact binaries in this phase and therefore constrain the effects of ET.
\par 
Still, we observe quantitative differences between the energy transferring and non energy transferring models during the final, inverted MT phase, which is long-lived and therefore more likely to be observed.
In particular, from the $q(t)$ curves in the bottom panel of Fig, \ref{fig:mdot}, we see that the ET model spends more time at mass ratios $q \gtrsim 1.4$ versus the no ET model. 
Furthermore, the no ET model monotonically approaches a mass ratio of unity without further inverting the mass ratio or the MT direction.
In contrast, the ET model experiences two inversions of the mass ratio at around $\tau=3.2$ and $\SI{4.0}{\mega yr}$ and the MT is inverted at around $\tau=\SI{3.3}{\mega yr}$.
\par
\begin{figure}
    \centering
    \includegraphics{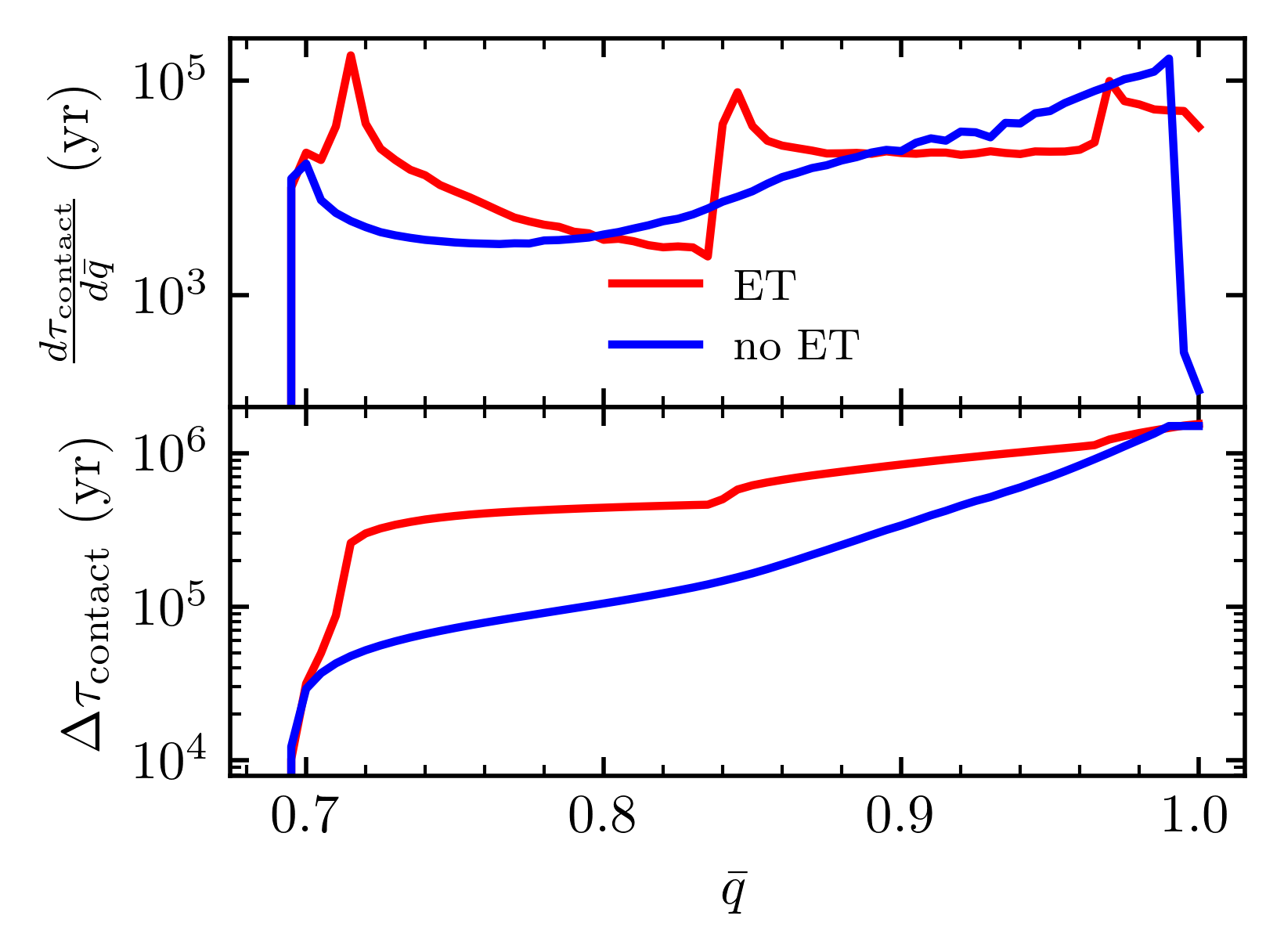}
    \caption{Top panel: Differential duration of contact (in bins of $d\bar{q}=0.005$) as function of observed mass ratio $\bar{q} = \min\left(\frac{M_1}{M_2}, \frac{M_2}{M_1}\right)$. Bottom panel: Cumulative duration of contact in configurations more extreme than $\bar{q}$.}
    \label{fig:dtdq}
\end{figure}

In Fig. \ref{fig:dtdq}, in the top panel we show the differential duration of the long-lived contact phase, plotted against $\bar{q} = \min\left(\frac{M_1}{M_2}, \frac{M_2}{M_1}\right)$\footnote{We use $\bar{q}$ here since in an observed system, one has no a priori information on what the most massive component is.}, while the cumulative duration is shown in the bottom panel.
The top panel shows three distinct peaks in the red $\bar{q}$ distribution, two of which correspond to the periods where the ET model is inverting its MT direction (at $q\approx1.4$ and $q\approx0.85$) and the last corresponds to a mass ratio close unity $\bar{q}\approx0.97$ at the end of the simulation.
Comparing the two former peaks to the no ET curve at those $\bar{q}$ means that the ET model spends more time at these mass ratios than the no ET model.
As a cumulative distribution, from the bottom panel of Fig. \ref{fig:dtdq}, we deduce that it is about four times more likely to observe the ET model in a configuration more extreme than $\bar{q} = 0.8$ than the no ET model, even though the total length of the contact phase is similar to within \SI{0.05}{\mega yr}.

\par

\begin{figure}
    \centering 
    \includegraphics{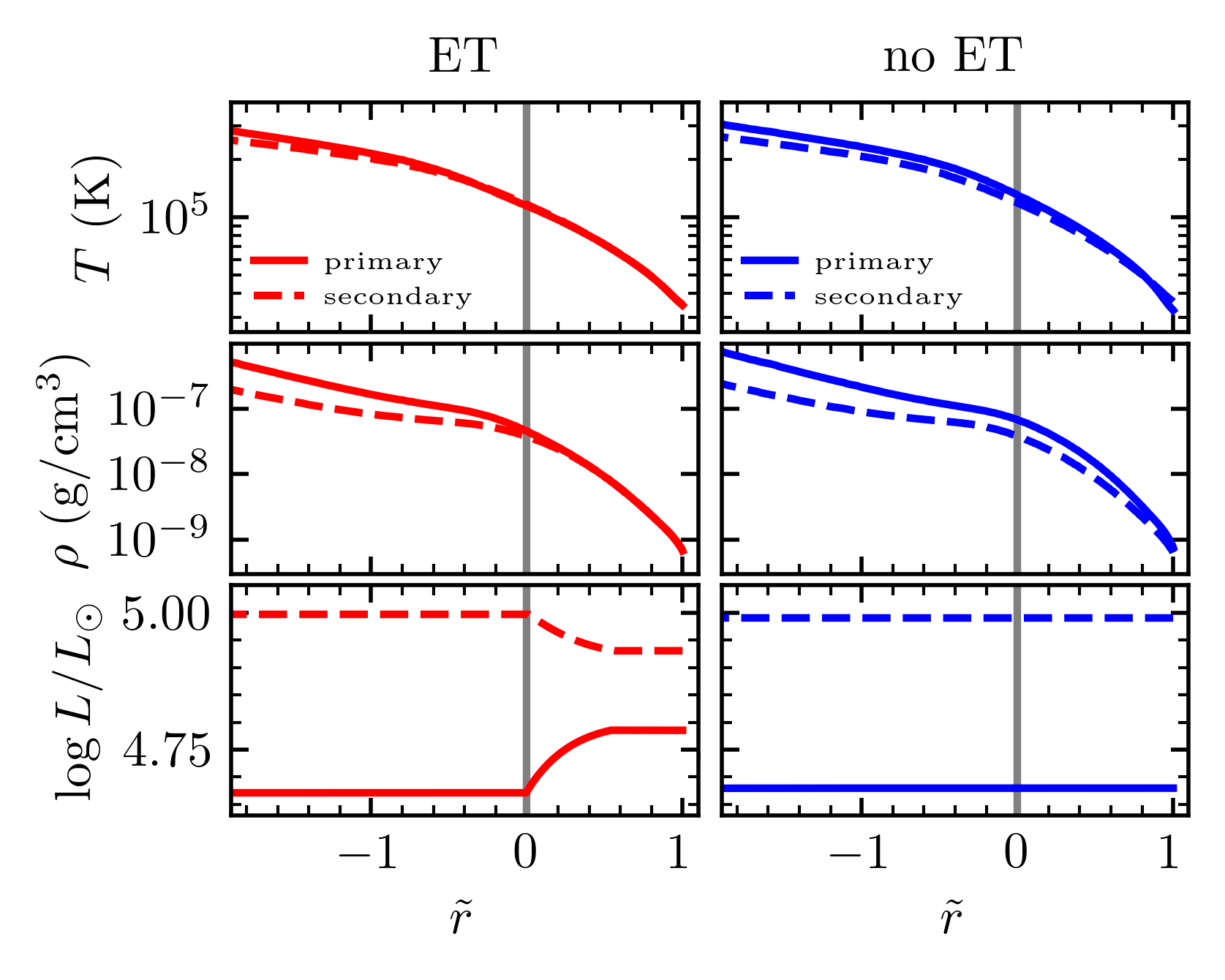}
    \caption{Profiles of outer layers of the binary components at the onset of inverse MT. It shows (top to bottom) the temperature, density and luminosity profiles for both components in the ET (red, left column) and no ET (blue, right column) cases, all as function of the scaled radius coordinate $\tilde{r} = \frac{r - R_{\rm RL}}{R  - R_{\rm RL}}$. The gray vertical lines show the location of the RL.}
    \label{fig:profiles}
\end{figure}

When comparing the profiles of the components in both models, we note a considerable difference in the structure of the outer layers of the components.
In Fig. \ref{fig:profiles}, the temperature, density and luminosity profiles of the models at the start of inverted MT phase ($\tau=\SI{2.7}{\mega yr}$) are shown.
The independent coordinate is chosen to be $\tilde{r} = \frac{r - R_{\rm RL}}{R  - R_{\rm RL}}$, so as to map the overflowing layers to the interval [0, 1].
At this point in the evolution, in both cases of ET and no ET, the stars are in a contact configuration and the primary is stripped to $M_1 = 18.3M_\odot$, while the secondary accreted mass to $M_2 = 26.4M_\odot$.
Note that these masses are close to the component masses \citet{abdul-masihConstrainingOvercontactPhase2021} found for V382 Cyg.
We see that in the system without ET, the overflowing shells do not satisfy our definition of shellularity since the temperature and density profiles of the primary and secondary do not match for constant Roche potential (which approximately corresponds to constant $\tilde{r}$). 
The ET models however match significantly better, as seen in the left column of Fig. \ref{fig:profiles} by the joining of the temperature and density profiles of the primary and secondary.
In particular, at the surface, the density of the ET components agree to within $0.2\%$, while the temperature to within $0.5\%$, whereas the surface properties of the no ET components vary more than $10\%$ (as expected for models of differing mass).
\par

An important shortcoming of our model is the assumed thickness of the ET layer.
\citet{shuStructureContactBinaries1979} argue that the thickness $d$ of the ET layer is on the order of:
\begin{equation}\label{eq:etthickness}
    \frac{d}{a} \sim \delta^{0.4},\quad \delta = \frac{L_{\rm trans}/a^2}{\rho h c_s}
\end{equation}
with $a$ the binary separation, and $\rho$, $h$ and $c_s$ the density, specific enthalpy and local sound speed evaluated at the RL, respectively.
\citet{lubowStructureContactBinaries1979} compute this number to be $d/a \sim 10^{-2}$ for stars of masses around 4-8$M_\odot$ which justifies that \citetalias{shuStructureContactBinaries1976} modeled the layer as a discontinuity in the stellar profile, located at the RL radius.
\par
\begin{figure}
    \centering
    \includegraphics{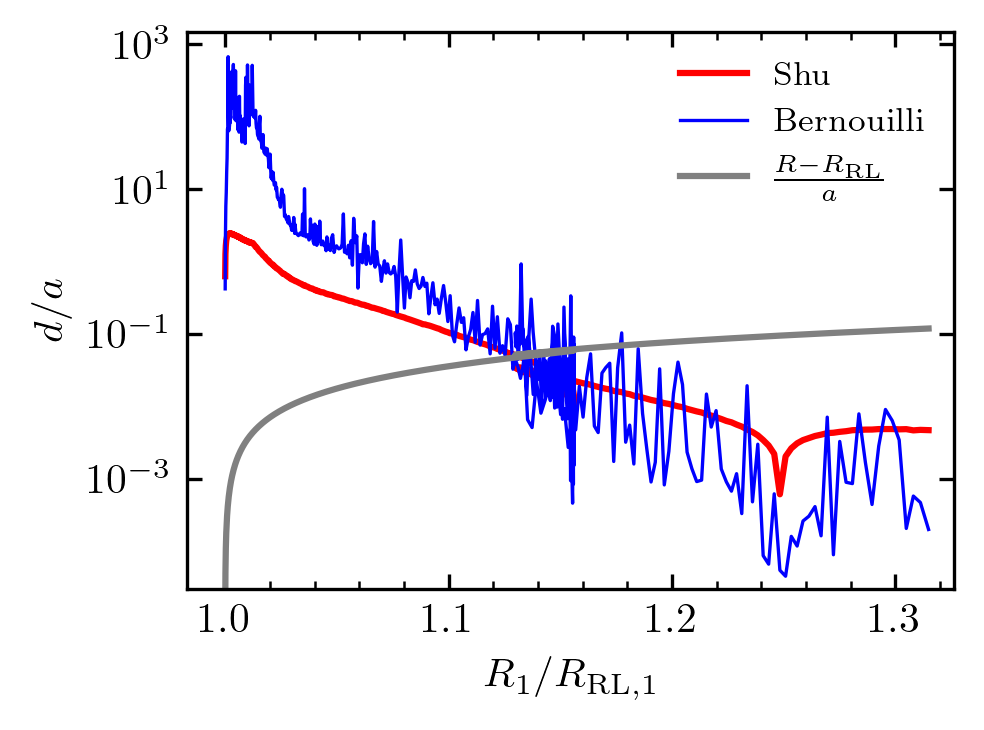}
    \caption{Thickness of the ET layer $d$ with respect to the binary separation $a$ as function of the radius of the primary star during the long lived contact phase. The gray line gives the physical size of the overflowing layers, and corresponds to the maximal width the ET layer can assume.}
    \label{fig:deltashu}
\end{figure}

However, direct computation of Eq. \eqref{eq:etthickness} for our $25 + 20 M_\odot$ model show that this estimation breaks down for higher masses, see the red line in Fig. \ref{fig:deltashu}.
This suggests that the energy redistribution flow, modeled as a discontinuity by \citetalias{shuStructureContactBinaries1976}, is not sufficient in these higher mass stars.
Except when the binary has reached considerable overflow, where $d/a \lesssim 10^{-2}$, we find that the thickness needed can be a significant fraction of the binary separation, even surpassing it in the early stages of the contact phase.
\par
Another way to compute an upper limit on the thickness of the ET layer is to use Bernouilli's equation.
If we consider fluid motion from far away from ${\rm L}_1$ on the primary star (location $i$) toward ${\rm L}_1$ (location $f$), we have
\begin{equation}
    \frac{1}{2}v_f^2 + \int_i^f\frac{dP}{\rho} = 0,
\end{equation}
where we have already canceled the potential terms $\Psi_i, \Psi_f$ since we move along an equipotential surface, and the initial velocity $v_i$ is assumed to be negligible.
Making then the estimation:
\begin{equation}
    \int_i^f\frac{dP}{\rho} \approx \left(\frac{1}{\rho_f}+\frac{1}{\rho_i}\right)\left(P_f-P_i\right),
\end{equation}
and assuming that the transferred energy through the binary neck of width $b$ by a mass flow $\dot{M} = \rho_i v_f b^2$ is
\begin{equation}
    L_{\rm trans} = \rho_i v_f b^2 (c_{p, 1} + c_{p, 2}) (T_f - T_i),
\end{equation}
we compute for the minimal thickness of the ET layer:
\begin{equation}\label{eq:width}
    b \approx \sqrt{\frac{L_{\rm trans}}{\rho_i (c_{p, 1} + c_{p, 2}) (T_f-T_i) \sqrt{\left(\frac{1}{\rho_f}+\frac{1}{\rho_i}\right)\left(P_f-P_i\right)}}}.
\end{equation}
Finally the thickness of the layer at the neck is related to the thickness far away from ${\rm L}_1$ by
\begin{equation}\label{eq:thickbernoui}
    \frac{d}{a} \approx \left(\frac{b}{a}\right)^2,
\end{equation}
since the Roche potential varies quadratically near ${\rm L}_1$ and linearly elsewhere.
Equation \eqref{eq:width} gives a lower limit on the width of ET layer so that a balanced mass flow $\dot{M}$ in the contact binary can carry a to be transferred luminosity $L_{\rm trans}$.
Conversely, it can be interpreted as $L_{\rm trans}$ being the maximal luminosity the mass flow can carry in a layer of fixed width $b$. 
\par
We plot also the Bernouilli computed thickness of Eq. \eqref{eq:thickbernoui} in Fig. \ref{fig:deltashu}.
Similarly, we see that the required thickness $d/a$ is much larger than what the radius of the primary star allows room for.
Only at later times, when the mass ratio has equilibrated and deeper contact is engaged is the estimated width smaller than the overflow rate of $R-R_{\rm RL}$ of the primary.

\par

\subsection{Mass vs luminosity ratios}
\begin{figure}
    \centering
    \includegraphics{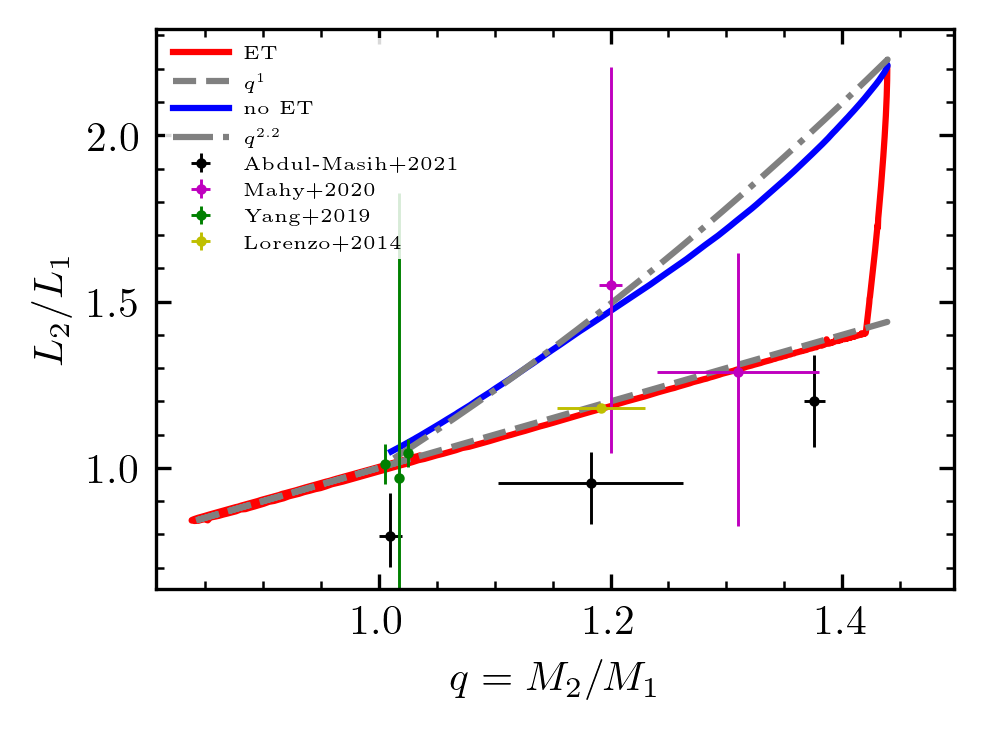}
    \caption{Luminosity ratio versus mass ratio during the nuclear timescale MT of the $19 + 14 M_\odot$ systems of Sect. \ref{ssec:detailed}.
    Overplotted are measurements from observed massive (near-)contact systems from \citet{abdul-masihConstrainingOvercontactPhase2021}, \citet{yangComprehensiveStudyThree2019} and \citet{lorenzoMYCamelopardalisVery2014}. The systems from \citet{mahyTarantulaMassiveBinary2020} were classified as `uncertain configurations'.}
    \label{fig:ML}
\end{figure}

During the nuclear timescale, inverted MT phase, contact is engaged so that our ET scheme acts to move luminosity from one component to the other. 
As mentioned in Sect. \ref{sec:ettheory}, we expect the luminosity ratio of contact binaries to follow the mass ratio $L \propto M$, as opposed to detached stars following a single star mass-luminosity relation $L \propto M^\alpha$ with $\alpha \simeq \numrange{2}{3}$.
Figure \ref{fig:ML} shows the evolution of the luminosity ratio as function of the mass ratio during the slow MT phase of the $25 + 20 M_\odot$ explored in Sect. \ref{ssec:detailed}.
In this graph, the models evolve from the top right at $q\approx 1.4$ to near equal mass ratio on the left.
We see that the model not including ET follow closely a $q^{2.2}$ relation, appropriate for single stars in the mass range of $\numrange{10}{30}M_\odot$.
As the models including ET engage into deep contact however, their mass-luminosity ratio changes drastically from the $q^{2.2}$ line in near contact to the $q^1$ relation in full contact.
\par
Overplotted on Fig. \ref{fig:ML} are measurements of several observed massive contact binaries of \citet{abdul-masihConstrainingOvercontactPhase2021, yangComprehensiveStudyThree2019} and \citet{lorenzoMYCamelopardalisVery2014} \citep[see also Fig. 2 of][]{langerOpenQuestionsMassive2022}.
Curiously, the luminosity ratios from \citet{abdul-masihConstrainingOvercontactPhase2021}, although following an $L \propto M$ trend, are offset to lower $L_2/L_1$ than predicted.
Either the luminosity of the primary is overestimated, or the uncertainties are underestimated.
The systems included from \citet{mahyTarantulaMassiveBinary2020} were categorized as `uncertain configurations' since the measurement of the radius was consistent with being above as well as below the RL.
In context of the mass-luminosity relation however, we expect that the system from \citet{mahyTarantulaMassiveBinary2020} at $q\approx 1.3$ (VFTS 563) is a true contact system, while the one at $q\approx 1.2$ (VFTS 217) is not (although within 1$\sigma$ it could be either).

\section{Conclusions}\label{sec:conc}
In this work we have taken a step forward in the detailed modeling of contact binaries, by implementing a model of ET in detailed stellar structure and evolution models.
From Fig. \ref{fig:profiles}, we see that this relatively simple model (that is, the inclusion of a heat source or sink in the stellar model as proxy for the ET) is capable of shellularizing the common layers of stellar components in a contact configuration.
Models without such ET do not exhibit their common layers to be shellular, which, from theoretical arguments, would drive strong horizontal flows equilibrating all gradients, in particular pressure.
\par

From Figs. \ref{fig:mdot} and \ref{fig:dtdq}, we saw that the time spent in deep contact at mass ratios between $\bar{q}=\numrange{0.7}{0.8}$ is extended if ET was included, versus when it was ignored.
This is a promising result, in that if this trend persists across the parameter space of total mass, initial mass ratio and initial period, ET could provide an answer to the discrepancy between the observed mass ratio distribution of massive contact systems and its predicted distribution.
The computation of a full grid of models and performing population synthesis is the topic of future work.
\par

\begin{acknowledgements}
M.F. thanks the Flemish research foundation (FWO, Fonds voor Wetenschappelijk Onderzoek) PhD fellowship No. 11H2421N for its support.
P.M. acknowledges support from the FWO junior postdoctoral fellowship No. 12ZY520N and the senior postdoctoral fellowship No. 12ZY523N.
The research leading to these results has received funding from the European Research Council (ERC) under the European Union's Horizon 2020 research and innovation program (grant agreement numbers 772225: MULTIPLES).
\end{acknowledgements}

%---------------------------------------------------------------------
\bibliographystyle{bibtex/aa}
\bibliography{bibtex/Massive_Stars.bib, bibtex/others}

\begin{thebibliography}{84}
\expandafter\ifx\csname natexlab\endcsname\relax\def\natexlab#1{#1}\fi

\bibitem[{{Abdul-Masih} {et~al.}(2021){Abdul-Masih}, Sana, Hawcroft, Almeida,
  Brands, {de Mink}, Justham, Langer, Mahy, Marchant, Menon, Puls, \&
  Sundqvist}]{abdul-masihConstrainingOvercontactPhase2021}
{Abdul-Masih}, M., Sana, H., Hawcroft, C., {et~al.} 2021, A\&A, 651, A96
  \csname abdul-masihConstrainingOvercontactPhase2021link\endcsname~\csname
  abdul-masihConstrainingOvercontactPhase2021note\endcsname

\bibitem[{{Aguilera-Dena} {et~al.}(2020){Aguilera-Dena}, Langer, Antoniadis, \&
  M{\"u}ller}]{aguilera-denaPrecollapsePropertiesSuperluminous2020}
{Aguilera-Dena}, D.~R., Langer, N., Antoniadis, J., \& M{\"u}ller, B. 2020,
  ApJ, 901, 114 \csname
  aguilera-denaPrecollapsePropertiesSuperluminous2020link\endcsname~\csname
  aguilera-denaPrecollapsePropertiesSuperluminous2020note\endcsname

\bibitem[{{Aguilera-Dena} {et~al.}(2018){Aguilera-Dena}, Langer, Moriya, \&
  Schootemeijer}]{aguilera-denaRelatedProgenitorModels2018}
{Aguilera-Dena}, D.~R., Langer, N., Moriya, T.~J., \& Schootemeijer, A. 2018,
  ApJ, 858, 115 \csname
  aguilera-denaRelatedProgenitorModels2018link\endcsname~\csname
  aguilera-denaRelatedProgenitorModels2018note\endcsname

\bibitem[{Alcock {et~al.}(1997)Alcock, Allsman, Alves, Axelrod, Becker,
  Bennett, Cook, Freeman, Griest, Lacy, Lehner, Marshall, Minniti, Peterson,
  Pratt, Quinn, Rodgers, Stubbs, Sutherland, \&
  Welch}]{alcockMACHOProjectLMC1997}
Alcock, C., Allsman, R.~A., Alves, D., {et~al.} 1997, Astron. J., 114, 326
  \csname alcockMACHOProjectLMC1997link\endcsname~\csname
  alcockMACHOProjectLMC1997note\endcsname

\bibitem[{Asplund {et~al.}(2009)Asplund, Grevesse, Sauval, \&
  Scott}]{asplundChemicalCompositionSun2009}
Asplund, M., Grevesse, N., Sauval, A.~J., \& Scott, P. 2009, ARAA, 47, 481
  \csname asplundChemicalCompositionSun2009link\endcsname~\csname
  asplundChemicalCompositionSun2009note\endcsname

\bibitem[{Biermann \& Thomas(1972)}]{biermannModelsContactBinaries1972}
Biermann, P. \& Thomas, H.~C. 1972, Astron. Astrophys., 16, 60 \csname
  biermannModelsContactBinaries1972link\endcsname~\csname
  biermannModelsContactBinaries1972note\endcsname

\bibitem[{Binnendijk(1970)}]{binnendijkOrbitalElementsUrsae1970}
Binnendijk, L. 1970, Vistas in Astronomy, 12, 217 \csname
  binnendijkOrbitalElementsUrsae1970link\endcsname~\csname
  binnendijkOrbitalElementsUrsae1970note\endcsname

\bibitem[{Blouin {et~al.}(2020)Blouin, Shaffer, Saumon, \&
  Starrett}]{blouinNewConductiveOpacities2020}
Blouin, S., Shaffer, N.~R., Saumon, D., \& Starrett, C.~E. 2020, ApJ, 899, 46
  \csname blouinNewConductiveOpacities2020link\endcsname~\csname
  blouinNewConductiveOpacities2020note\endcsname

\bibitem[{{B{\"o}hm-Vitense}(1958)}]{bohm-vitenseUberWasserstoffkonvektionszoneSternen1958}
{B{\"o}hm-Vitense}, E. 1958, Z. Astrophys., 46, 108 \csname
  bohm-vitenseUberWasserstoffkonvektionszoneSternen1958link\endcsname~\csname
  bohm-vitenseUberWasserstoffkonvektionszoneSternen1958note\endcsname

\bibitem[{Bondi \& Hoyle(1944)}]{bondiMechanismAccretionStars1944}
Bondi, H. \& Hoyle, F. 1944, Mon. Not. R. Astron. Soc., 104, 273 \csname
  bondiMechanismAccretionStars1944link\endcsname~\csname
  bondiMechanismAccretionStars1944note\endcsname

\bibitem[{Brott {et~al.}(2011)Brott, de~Mink, Cantiello, Langer, de~Koter,
  Evans, Hunter, Trundle, \& Vink}]{brottRotatingMassiveMainsequence2011}
Brott, I., de~Mink, S.~E., Cantiello, M., {et~al.} 2011, A\&A, 530, A115
  \csname brottRotatingMassiveMainsequence2011link\endcsname~\csname
  brottRotatingMassiveMainsequence2011note\endcsname

\bibitem[{Cassisi {et~al.}(2007)Cassisi, Potekhin, Pietrinferni, Catelan, \&
  Salaris}]{cassisiUpdatedElectronconductionOpacities2007}
Cassisi, S., Potekhin, A.~Y., Pietrinferni, A., Catelan, M., \& Salaris, M.
  2007, ApJ, 661, 1094 \csname
  cassisiUpdatedElectronconductionOpacities2007link\endcsname~\csname
  cassisiUpdatedElectronconductionOpacities2007note\endcsname

\bibitem[{Chugunov {et~al.}(2007)Chugunov, Dewitt, \&
  Yakovlev}]{chugunovCoulombTunnelingFusion2007}
Chugunov, A.~I., Dewitt, H.~E., \& Yakovlev, D.~G. 2007, PhysRevD, 76, 025028
  \csname chugunovCoulombTunnelingFusion2007link\endcsname~\csname
  chugunovCoulombTunnelingFusion2007note\endcsname

\bibitem[{Cox \& Giuli(1968)}]{coxPrinciplesStellarStructure1968}
Cox, J.~P. \& Giuli, R.~T. 1968, Principles of Stellar Structure ({New York}:
  {Gordon and Breach}) \csname
  coxPrinciplesStellarStructure1968link\endcsname~\csname
  coxPrinciplesStellarStructure1968note\endcsname

\bibitem[{Cyburt {et~al.}(2010)Cyburt, Amthor, Ferguson, Meisel, Smith, Warren,
  Heger, Hoffman, Rauscher, Sakharuk, Schatz, Thielemann, \&
  Wiescher}]{cyburtJINAREACLIBDatabase2010}
Cyburt, R.~H., Amthor, A.~M., Ferguson, R., {et~al.} 2010, Astrophys. J. Suppl.
  Ser., 189, 240 \csname cyburtJINAREACLIBDatabase2010link\endcsname~\csname
  cyburtJINAREACLIBDatabase2010note\endcsname

\bibitem[{{de Mink} {et~al.}(2014){de Mink}, Sana, Langer, Izzard, \&
  Schneider}]{deminkIncidenceStellarMergers2014}
{de Mink}, S.~E., Sana, H., Langer, N., Izzard, R.~G., \& Schneider, F. R.~N.
  2014, ApJ, 782, 7 \csname
  deminkIncidenceStellarMergers2014link\endcsname~\csname
  deminkIncidenceStellarMergers2014note\endcsname

\bibitem[{Eggen(1967)}]{eggenContactBinariesII1967}
Eggen, O.~J. 1967, Mem. R. Astron. Soc., 70, 111 \csname
  eggenContactBinariesII1967link\endcsname~\csname
  eggenContactBinariesII1967note\endcsname

\bibitem[{Evans {et~al.}(2011)Evans, Taylor, {H{\'e}nault-Brunet}, Sana, {de
  Koter}, {Sim{\'o}n-D{\'i}az}, Carraro, Bagnoli, Bastian, Bestenlehner,
  Bonanos, Bressert, Brott, Campbell, Cantiello, Clark, Costa, Crowther, {de
  Mink}, Doran, Dufton, Dunstall, Friedrich, Garcia, Gieles, Gr{\"a}fener,
  Herrero, Howarth, Izzard, Langer, Lennon, Ma{\'i}z~Apell{\'a}niz, Markova,
  Najarro, Puls, Ramirez, {Sab{\'i}n-Sanjuli{\'a}n}, Smartt, Stroud, {van
  Loon}, Vink, \& Walborn}]{evansVLTFLAMESTarantulaSurvey2011}
Evans, C.~J., Taylor, W.~D., {H{\'e}nault-Brunet}, V., {et~al.} 2011, A\&A,
  530, A108 \csname evansVLTFLAMESTarantulaSurvey2011link\endcsname~\csname
  evansVLTFLAMESTarantulaSurvey2011note\endcsname

\bibitem[{Fabry {et~al.}(2022)Fabry, Marchant, \&
  Sana}]{fabryModelingOvercontactBinaries2022}
Fabry, M., Marchant, P., \& Sana, H. 2022, A\&A, 661, A123 \csname
  fabryModelingOvercontactBinaries2022link\endcsname~\csname
  fabryModelingOvercontactBinaries2022note\endcsname

\bibitem[{Ferguson {et~al.}(2005)Ferguson, Alexander, Allard, Barman, Bodnarik,
  Hauschildt, {Heffner-Wong}, \& Tamanai}]{fergusonLowTemperatureOpacities2005}
Ferguson, J.~W., Alexander, D.~R., Allard, F., {et~al.} 2005, Astrophys. J.,
  623, 585 \csname fergusonLowTemperatureOpacities2005link\endcsname~\csname
  fergusonLowTemperatureOpacities2005note\endcsname

\bibitem[{Ferrario {et~al.}(2009)Ferrario, Pringle, Tout, \&
  Wickramasinghe}]{ferrarioOriginMagnetismUpper2009}
Ferrario, L., Pringle, J.~E., Tout, C.~A., \& Wickramasinghe, D.~T. 2009,
  MNRAS, 400, L71 \csname
  ferrarioOriginMagnetismUpper2009link\endcsname~\csname
  ferrarioOriginMagnetismUpper2009note\endcsname

\bibitem[{Flannery(1976)}]{flanneryCyclicThermalInstability1976}
Flannery, B.~P. 1976, Astrophys. J., 205, 217 \csname
  flanneryCyclicThermalInstability1976link\endcsname~\csname
  flanneryCyclicThermalInstability1976note\endcsname

\bibitem[{Fossati {et~al.}(2015)Fossati, Castro, Sch{\"o}ller, Hubrig, Langer,
  Morel, Briquet, Herrero, Przybilla, Sana, Schneider, {de Koter}, \& {BOB
  Collaboration}}]{fossatiFieldsOBStars2015}
Fossati, L., Castro, N., Sch{\"o}ller, M., {et~al.} 2015, Astron. Astrophys.,
  582, A45 \csname fossatiFieldsOBStars2015link\endcsname~\csname
  fossatiFieldsOBStars2015note\endcsname

\bibitem[{Frost {et~al.}(2023)Frost, Sana, Mahy, Wade, Barron, Bouquin,
  M´erand, Schneider, Shenar, Barb´a†, Bowman, Fabry, Farhang, Marchant,
  Morrell, \& Smoker}]{frost2023}
Frost, A.~J., Sana, H., Mahy, L., {et~al.} 2023, Science, subm. \csname
  frost2023link\endcsname~\csname frost2023note\endcsname

\bibitem[{Fuller {et~al.}(1985)Fuller, Fowler, \&
  Newman}]{fullerStellarWeakInteraction1985}
Fuller, G.~M., Fowler, W.~A., \& Newman, M.~J. 1985, ApJ, 293, 1 \csname
  fullerStellarWeakInteraction1985link\endcsname~\csname
  fullerStellarWeakInteraction1985note\endcsname

\bibitem[{Gr{\"a}fener {et~al.}(2011)Gr{\"a}fener, Vink, {de Koter}, \&
  Langer}]{grafenerEddingtonFactorKey2011}
Gr{\"a}fener, G., Vink, J.~S., {de Koter}, A., \& Langer, N. 2011, A\&A, 535,
  A56 \csname grafenerEddingtonFactorKey2011link\endcsname~\csname
  grafenerEddingtonFactorKey2011note\endcsname

\bibitem[{Grunhut {et~al.}(2017)Grunhut, Wade, Neiner, Oksala, Petit, Alecian,
  Bohlender, Bouret, Henrichs, Hussain, Kochukhov, \& {MiMeS
  Collaboration}}]{grunhutMiMeSSurveyMagnetism2017b}
Grunhut, J.~H., Wade, G.~A., Neiner, C., {et~al.} 2017, Mon. Not. R. Astron.
  Soc., 465, 2432 \csname
  grunhutMiMeSSurveyMagnetism2017blink\endcsname~\csname
  grunhutMiMeSSurveyMagnetism2017bnote\endcsname

\bibitem[{Hamann {et~al.}(1995)Hamann, Koesterke, \&
  Wessolowski}]{hamannSpectralAnalysesGalactic1995}
Hamann, W.-R., Koesterke, L., \& Wessolowski, U. 1995, Astron. Astrophys. V299
  P151, 299, 151 \csname
  hamannSpectralAnalysesGalactic1995link\endcsname~\csname
  hamannSpectralAnalysesGalactic1995note\endcsname

\bibitem[{Hazlehurst(1985)}]{hazlehurstDissipationFactorContact1985}
Hazlehurst, J. 1985, Astron. Astrophys., 145, 25 \csname
  hazlehurstDissipationFactorContact1985link\endcsname~\csname
  hazlehurstDissipationFactorContact1985note\endcsname

\bibitem[{Hazlehurst(1993)}]{hazlehurstEquilibriumContactBinary1993}
Hazlehurst, J. 1993, Astron. Astrophys. Vol 271 P 209 1993, 271, 209 \csname
  hazlehurstEquilibriumContactBinary1993link\endcsname~\csname
  hazlehurstEquilibriumContactBinary1993note\endcsname

\bibitem[{Hurley {et~al.}(2002)Hurley, Tout, \&
  Pols}]{hurleyEvolutionBinaryStars2002}
Hurley, J.~R., Tout, C.~A., \& Pols, O.~R. 2002, Mon. Not. R. Astron. Soc.,
  329, 897 \csname hurleyEvolutionBinaryStars2002link\endcsname~\csname
  hurleyEvolutionBinaryStars2002note\endcsname

\bibitem[{Iglesias \& Rogers(1993)}]{iglesiasRadiativeOpacitiesCarbon1993}
Iglesias, C.~A. \& Rogers, F.~J. 1993, ApJ, 412, 752 \csname
  iglesiasRadiativeOpacitiesCarbon1993link\endcsname~\csname
  iglesiasRadiativeOpacitiesCarbon1993note\endcsname

\bibitem[{Iglesias \& Rogers(1996)}]{iglesiasUpdatedOpalOpacities1996}
Iglesias, C.~A. \& Rogers, F.~J. 1996, ApJ, 464, 943 \csname
  iglesiasUpdatedOpalOpacities1996link\endcsname~\csname
  iglesiasUpdatedOpalOpacities1996note\endcsname

\bibitem[{Itoh {et~al.}(1996)Itoh, Hayashi, Nishikawa, \&
  Kohyama}]{itohNeutrinoEnergyLoss1996}
Itoh, N., Hayashi, H., Nishikawa, A., \& Kohyama, Y. 1996, ApJS, 102, 411
  \csname itohNeutrinoEnergyLoss1996link\endcsname~\csname
  itohNeutrinoEnergyLoss1996note\endcsname

\bibitem[{Jermyn {et~al.}(2023)Jermyn, Bauer, Schwab, Farmer, Ball, Bellinger,
  Dotter, Joyce, Marchant, Mombarg, Wolf, Sunny~Wong, Cinquegrana, Farrell,
  Smolec, Thoul, Cantiello, Herwig, Toloza, Bildsten, Townsend, \&
  Timmes}]{jermynModulesExperimentsStellar2023}
Jermyn, A.~S., Bauer, E.~B., Schwab, J., {et~al.} 2023, AJSS, 265, 15 \csname
  jermynModulesExperimentsStellar2023link\endcsname~\csname
  jermynModulesExperimentsStellar2023note\endcsname

\bibitem[{Jermyn {et~al.}(2021)Jermyn, Schwab, Bauer, Timmes, \&
  Potekhin}]{jermynSkyeDifferentiableEquation2021}
Jermyn, A.~S., Schwab, J., Bauer, E., Timmes, F.~X., \& Potekhin, A.~Y. 2021,
  ApJ, 913, 72 \csname
  jermynSkyeDifferentiableEquation2021link\endcsname~\csname
  jermynSkyeDifferentiableEquation2021note\endcsname

\bibitem[{K{\"a}hler(1989)}]{kahlerStructureEquationsContact1989}
K{\"a}hler, H. 1989, Astron. Astrophys., 209, 67 \csname
  kahlerStructureEquationsContact1989link\endcsname~\csname
  kahlerStructureEquationsContact1989note\endcsname

\bibitem[{K{\"a}hler(2004)}]{kahlerStructureContactBinaries2004}
K{\"a}hler, H. 2004, A\&A, 414, 317 \csname
  kahlerStructureContactBinaries2004link\endcsname~\csname
  kahlerStructureContactBinaries2004note\endcsname

\bibitem[{Kippenhahn {et~al.}(1980)Kippenhahn, Ruschenplatt, \&
  Thomas}]{kippenhahnTimeScaleThermohaline1980}
Kippenhahn, R., Ruschenplatt, G., \& Thomas, H.-C. 1980, A\&A, 91, 175 \csname
  kippenhahnTimeScaleThermohaline1980link\endcsname~\csname
  kippenhahnTimeScaleThermohaline1980note\endcsname

\bibitem[{K{\"o}hler {et~al.}(2015)K{\"o}hler, Langer, {de Koter}, {de Mink},
  Crowther, Evans, Gr{\"a}fener, Sana, Sanyal, Schneider, \&
  Vink}]{kohlerEvolutionRotatingVery2015}
K{\"o}hler, K., Langer, N., {de Koter}, A., {et~al.} 2015, Astron. Astrophys.,
  573, A71 \csname kohlerEvolutionRotatingVery2015link\endcsname~\csname
  kohlerEvolutionRotatingVery2015note\endcsname

\bibitem[{Kuiper(1941)}]{kuiperInterpretationLyraeOther1941}
Kuiper, G.~P. 1941, ApJ, 93, 133 \csname
  kuiperInterpretationLyraeOther1941link\endcsname~\csname
  kuiperInterpretationLyraeOther1941note\endcsname

\bibitem[{Langanke \&
  {Mart{\'i}nez-Pinedo}(2000)}]{langankeShellmodelCalculationsStellar2000}
Langanke, K. \& {Mart{\'i}nez-Pinedo}, G. 2000, Nucl. Phys. A, 673, 481 \csname
  langankeShellmodelCalculationsStellar2000link\endcsname~\csname
  langankeShellmodelCalculationsStellar2000note\endcsname

\bibitem[{Langer(2022)}]{langerOpenQuestionsMassive2022}
Langer, N. 2022, arXiv e-prints, 2209.04165 \csname
  langerOpenQuestionsMassive2022link\endcsname~\csname
  langerOpenQuestionsMassive2022note\endcsname

\bibitem[{Langer {et~al.}(1983)Langer, Fricke, \&
  Sugimoto}]{langerSemiconvectiveDiffusionEnergy1983}
Langer, N., Fricke, K.~J., \& Sugimoto, D. 1983, A\&A, 126, 207 \csname
  langerSemiconvectiveDiffusionEnergy1983link\endcsname~\csname
  langerSemiconvectiveDiffusionEnergy1983note\endcsname

\bibitem[{Lorenzo {et~al.}(2014)Lorenzo, Negueruela, Baker, Garc{\'i}a,
  {Sim{\'o}n-D{\'i}az}, Pastor, \&
  M{\'e}ndez~Majuelos}]{lorenzoMYCamelopardalisVery2014}
Lorenzo, J., Negueruela, I., Baker, A. K. F.~V., {et~al.} 2014, Astron.
  Astrophys., 572, A110 \csname
  lorenzoMYCamelopardalisVery2014link\endcsname~\csname
  lorenzoMYCamelopardalisVery2014note\endcsname

\bibitem[{Lorenzo {et~al.}(2017)Lorenzo, {Sim{\'o}n-D{\'i}az}, Negueruela,
  Vilardell, Garcia, Evans, \& Montes}]{lorenzoMassiveMultipleSystem2017}
Lorenzo, J., {Sim{\'o}n-D{\'i}az}, S., Negueruela, I., {et~al.} 2017, Astron.
  Astrophys., 606, A54 \csname
  lorenzoMassiveMultipleSystem2017link\endcsname~\csname
  lorenzoMassiveMultipleSystem2017note\endcsname

\bibitem[{Lubow \& Shu(1977)}]{lubowStructureContactBinaries1977}
Lubow, S.~H. \& Shu, F.~H. 1977, Astrophys. J., 216, 517 \csname
  lubowStructureContactBinaries1977link\endcsname~\csname
  lubowStructureContactBinaries1977note\endcsname

\bibitem[{Lubow \& Shu(1979)}]{lubowStructureContactBinaries1979}
Lubow, S.~H. \& Shu, F.~H. 1979, Astrophys. J., 229, 657 \csname
  lubowStructureContactBinaries1979link\endcsname~\csname
  lubowStructureContactBinaries1979note\endcsname

\bibitem[{Lucy(1968)}]{lucyStructureContactBinaries1968}
Lucy, L.~B. 1968, ApJ, 151, 1123 \csname
  lucyStructureContactBinaries1968link\endcsname~\csname
  lucyStructureContactBinaries1968note\endcsname

\bibitem[{Lucy(1976)}]{lucyUrsaeMajorisSystems1976}
Lucy, L.~B. 1976, Astrophys. J., 205, 208 \csname
  lucyUrsaeMajorisSystems1976link\endcsname~\csname
  lucyUrsaeMajorisSystems1976note\endcsname

\bibitem[{Mahy {et~al.}(2020)Mahy, Almeida, Sana, Clark, {de Koter}, {de Mink},
  Evans, Grin, Langer, Moffat, Schneider, Shenar, \&
  Tramper}]{mahyTarantulaMassiveBinary2020}
Mahy, L., Almeida, L.~A., Sana, H., {et~al.} 2020, A\&A, 634, A119 \csname
  mahyTarantulaMassiveBinary2020link\endcsname~\csname
  mahyTarantulaMassiveBinary2020note\endcsname

\bibitem[{Mandel \& {de Mink}(2016)}]{mandelMergingBinaryBlack2016}
Mandel, I. \& {de Mink}, S.~E. 2016, Mon. Not. R. Astron. Soc., 458, 2634
  \csname mandelMergingBinaryBlack2016link\endcsname~\csname
  mandelMergingBinaryBlack2016note\endcsname

\bibitem[{Marchant(2016)}]{marchantImpactTidesMass2016}
Marchant, P. 2016, PhD thesis, Bonn \csname
  marchantImpactTidesMass2016link\endcsname~\csname
  marchantImpactTidesMass2016note\endcsname

\bibitem[{Marchant {et~al.}(2016)Marchant, Langer, Podsiadlowski, Tauris, \&
  Moriya}]{marchantNewRouteMerging2016}
Marchant, P., Langer, N., Podsiadlowski, P., Tauris, T.~M., \& Moriya, T.~J.
  2016, A\&A, 588, A50 \csname
  marchantNewRouteMerging2016link\endcsname~\csname
  marchantNewRouteMerging2016note\endcsname

\bibitem[{Marchant {et~al.}(2021)Marchant, Pappas, {Gallegos-Garcia}, Berry,
  Taam, Kalogera, \& Podsiadlowski}]{marchantRoleMassTransfer2021}
Marchant, P., Pappas, K. M.~W., {Gallegos-Garcia}, M., {et~al.} 2021, Astron.
  Astrophys., 650, A107 \csname
  marchantRoleMassTransfer2021link\endcsname~\csname
  marchantRoleMassTransfer2021note\endcsname

\bibitem[{Menon {et~al.}(2021)Menon, Langer, {de Mink}, Justham, Sen,
  Sz{\'e}csi, {de Koter}, {Abdul-Masih}, Sana, Mahy, \&
  Marchant}]{menonDetailedEvolutionaryModels2021}
Menon, A., Langer, N., {de Mink}, S.~E., {et~al.} 2021, MNRAS, 507, 5013
  \csname menonDetailedEvolutionaryModels2021link\endcsname~\csname
  menonDetailedEvolutionaryModels2021note\endcsname

\bibitem[{Nieuwenhuijzen \& {de
  Jager}(1995)}]{nieuwenhuijzenAtmosphericAccelerationsStability1995}
Nieuwenhuijzen, H. \& {de Jager}, C. 1995, A\&A, 302, 811 \csname
  nieuwenhuijzenAtmosphericAccelerationsStability1995link\endcsname~\csname
  nieuwenhuijzenAtmosphericAccelerationsStability1995note\endcsname

\bibitem[{Oda {et~al.}(1994)Oda, Hino, Muto, Takahara, \&
  Sato}]{odaRateTablesWeak1994}
Oda, T., Hino, M., Muto, K., Takahara, M., \& Sato, K. 1994, At. Data Nucl.
  Data Tables, 56, 231 \csname odaRateTablesWeak1994link\endcsname~\csname
  odaRateTablesWeak1994note\endcsname

\bibitem[{Paxton {et~al.}(2011)Paxton, Bildsten, Dotter, Herwig, Lesaffre, \&
  Timmes}]{paxtonModulesExperimentsStellar2011}
Paxton, B., Bildsten, L., Dotter, A., {et~al.} 2011, AJSS, 192, 3 \csname
  paxtonModulesExperimentsStellar2011link\endcsname~\csname
  paxtonModulesExperimentsStellar2011note\endcsname

\bibitem[{Paxton {et~al.}(2013)Paxton, Cantiello, Arras, Bildsten, Brown,
  Dotter, Mankovich, Montgomery, Stello, Timmes, \&
  Townsend}]{paxtonModulesExperimentsStellar2013}
Paxton, B., Cantiello, M., Arras, P., {et~al.} 2013, AJSS, 208, 4 \csname
  paxtonModulesExperimentsStellar2013link\endcsname~\csname
  paxtonModulesExperimentsStellar2013note\endcsname

\bibitem[{Paxton {et~al.}(2015)Paxton, Marchant, Schwab, Bauer, Bildsten,
  Cantiello, Dessart, Farmer, Hu, Langer, Townsend, Townsley, \&
  Timmes}]{paxtonModulesExperimentsStellar2015}
Paxton, B., Marchant, P., Schwab, J., {et~al.} 2015, AJSS, 220, 15 \csname
  paxtonModulesExperimentsStellar2015link\endcsname~\csname
  paxtonModulesExperimentsStellar2015note\endcsname

\bibitem[{Paxton {et~al.}(2018)Paxton, Schwab, Bauer, Bildsten, Blinnikov,
  Duffell, Farmer, Goldberg, Marchant, Sorokina, Thoul, Townsend, \&
  Timmes}]{paxtonModulesExperimentsStellar2018}
Paxton, B., Schwab, J., Bauer, E.~B., {et~al.} 2018, AJSS, 234, 34 \csname
  paxtonModulesExperimentsStellar2018link\endcsname~\csname
  paxtonModulesExperimentsStellar2018note\endcsname

\bibitem[{Paxton {et~al.}(2019)Paxton, Smolec, Schwab, Gautschy, Bildsten,
  Cantiello, Dotter, Farmer, Goldberg, Jermyn, Kanbur, Marchant, Thoul,
  Townsend, Wolf, Zhang, \& Timmes}]{paxtonModulesExperimentsStellar2019}
Paxton, B., Smolec, R., Schwab, J., {et~al.} 2019, AJSS, 243, 10 \csname
  paxtonModulesExperimentsStellar2019link\endcsname~\csname
  paxtonModulesExperimentsStellar2019note\endcsname

\bibitem[{Potekhin \& Chabrier(2010)}]{potekhinThermodynamicFunctionsDense2010}
Potekhin, A.~Y. \& Chabrier, G. 2010, Contrib. Plasma Phys., 50, 82 \csname
  potekhinThermodynamicFunctionsDense2010link\endcsname~\csname
  potekhinThermodynamicFunctionsDense2010note\endcsname

\bibitem[{Poutanen(2017)}]{poutanenRosselandFluxMean2017}
Poutanen, J. 2017, ApJ, 835, 119 \csname
  poutanenRosselandFluxMean2017link\endcsname~\csname
  poutanenRosselandFluxMean2017note\endcsname

\bibitem[{Rogers \& Nayfonov(2002)}]{rogersUpdatedExpandedOPAL2002}
Rogers, F.~J. \& Nayfonov, A. 2002, Astrophys. J., 576, 1064 \csname
  rogersUpdatedExpandedOPAL2002link\endcsname~\csname
  rogersUpdatedExpandedOPAL2002note\endcsname

\bibitem[{Sana {et~al.}(2012)Sana, de~Mink, de~Koter, Langer, Evans, Gieles,
  Gosset, Izzard, Bouquin, \& Schneider}]{sanaBinaryInteractionDominates2012}
Sana, H., de~Mink, S.~E., de~Koter, A., {et~al.} 2012, Sci, 337, 444 \csname
  sanaBinaryInteractionDominates2012link\endcsname~\csname
  sanaBinaryInteractionDominates2012note\endcsname

\bibitem[{Saumon {et~al.}(1995)Saumon, Chabrier, \& {van
  Horn}}]{saumonEquationStateLowMass1995}
Saumon, D., Chabrier, G., \& {van Horn}, H.~M. 1995, Astrophys. J. Suppl. Ser.,
  99, 713 \csname saumonEquationStateLowMass1995link\endcsname~\csname
  saumonEquationStateLowMass1995note\endcsname

\bibitem[{Schneider {et~al.}(2019)Schneider, Ohlmann, Podsiadlowski, R{\"o}pke,
  Balbus, Pakmor, \& Springel}]{schneiderStellarMergersOrigin2019}
Schneider, F. R.~N., Ohlmann, S.~T., Podsiadlowski, P., {et~al.} 2019, Natur,
  574, 211 \csname schneiderStellarMergersOrigin2019link\endcsname~\csname
  schneiderStellarMergersOrigin2019note\endcsname

\bibitem[{Schneider {et~al.}(2016)Schneider, Podsiadlowski, Langer, Castro, \&
  Fossati}]{schneiderRejuvenationStellarMergers2016}
Schneider, F. R.~N., Podsiadlowski, P., Langer, N., Castro, N., \& Fossati, L.
  2016, MNRAS, 457, 2355 \csname
  schneiderRejuvenationStellarMergers2016link\endcsname~\csname
  schneiderRejuvenationStellarMergers2016note\endcsname

\bibitem[{Schootemeijer {et~al.}(2019)Schootemeijer, Langer, Grin, \&
  Wang}]{schootemeijerConstrainingMixingMassive2019}
Schootemeijer, A., Langer, N., Grin, N.~J., \& Wang, C. 2019, A\&A, 625, A132
  \csname schootemeijerConstrainingMixingMassive2019link\endcsname~\csname
  schootemeijerConstrainingMixingMassive2019note\endcsname

\bibitem[{Sen {et~al.}(2022)Sen, Langer, Marchant, Menon, {de Mink},
  Schootemeijer, Sch{\"u}rmann, Mahy, Hastings, Nathaniel, Sana, Wang, \&
  Xu}]{senDetailedModelsInteracting2022}
Sen, K., Langer, N., Marchant, P., {et~al.} 2022, Astron. Astrophys., 659, A98
  \csname senDetailedModelsInteracting2022link\endcsname~\csname
  senDetailedModelsInteracting2022note\endcsname

\bibitem[{Shu {et~al.}(1976)Shu, Lubow, \&
  Anderson}]{shuStructureContactBinaries1976}
Shu, F.~H., Lubow, S.~H., \& Anderson, L. 1976, ApJ, 209, 536 \csname
  shuStructureContactBinaries1976link\endcsname~\csname
  shuStructureContactBinaries1976note\endcsname

\bibitem[{Shu {et~al.}(1979)Shu, Lubow, \&
  Anderson}]{shuStructureContactBinaries1979}
Shu, F.~H., Lubow, S.~H., \& Anderson, L. 1979, ApJ, 229, 223 \csname
  shuStructureContactBinaries1979link\endcsname~\csname
  shuStructureContactBinaries1979note\endcsname

\bibitem[{Szymanski {et~al.}(2001)Szymanski, Kubiak, \&
  Udalski}]{szymanskiContactBinariesOGLEI2001}
Szymanski, M., Kubiak, M., \& Udalski, A. 2001, Acta Astron., 51, 259 \csname
  szymanskiContactBinariesOGLEI2001link\endcsname~\csname
  szymanskiContactBinariesOGLEI2001note\endcsname

\bibitem[{Tassoul(2000)}]{tassoulStellarRotation2000}
Tassoul, J.-L. 2000, Stellar {{Rotation}} ({Cambridge University Press})
  \csname tassoulStellarRotation2000link\endcsname~\csname
  tassoulStellarRotation2000note\endcsname

\bibitem[{Timmes \& Swesty(2000)}]{timmesAccuracyConsistencySpeed2000}
Timmes, F.~X. \& Swesty, F.~D. 2000, Astrophys. J. Suppl. Ser., 126, 501
  \csname timmesAccuracyConsistencySpeed2000link\endcsname~\csname
  timmesAccuracyConsistencySpeed2000note\endcsname

\bibitem[{Vanbeveren {et~al.}(1998)Vanbeveren, De~Donder, Van~Bever,
  Van~Rensbergen, \& De~Loore}]{vanbeverenWROtypeStar1998}
Vanbeveren, D., De~Donder, E., Van~Bever, J., Van~Rensbergen, W., \& De~Loore,
  C. 1998, New Astronomy, 3, 443 \csname
  vanbeverenWROtypeStar1998link\endcsname~\csname
  vanbeverenWROtypeStar1998note\endcsname

\bibitem[{Vink {et~al.}(2001)Vink, {de Koter}, \&
  Lamers}]{vinkMasslossPredictionsStars2001}
Vink, J.~S., {de Koter}, A., \& Lamers, H. J. G. L.~M. 2001, A\&A, 369, 574
  \csname vinkMasslossPredictionsStars2001link\endcsname~\csname
  vinkMasslossPredictionsStars2001note\endcsname

\bibitem[{{von Zeipel}(1924)}]{vonzeipelRadiativeEquilibriumRotating1924}
{von Zeipel}, H. 1924, MNRAS, 84, 665 \csname
  vonzeipelRadiativeEquilibriumRotating1924link\endcsname~\csname
  vonzeipelRadiativeEquilibriumRotating1924note\endcsname

\bibitem[{Wickramasinghe {et~al.}(2014)Wickramasinghe, Tout, \&
  Ferrario}]{wickramasingheMostMagneticStars2014}
Wickramasinghe, D.~T., Tout, C.~A., \& Ferrario, L. 2014, MNRAS, 437, 675
  \csname wickramasingheMostMagneticStars2014link\endcsname~\csname
  wickramasingheMostMagneticStars2014note\endcsname

\bibitem[{Yang {et~al.}(2019)Yang, Yuan, \&
  Dai}]{yangComprehensiveStudyThree2019}
Yang, Y., Yuan, H., \& Dai, H. 2019, AJ, 157, 111 \csname
  yangComprehensiveStudyThree2019link\endcsname~\csname
  yangComprehensiveStudyThree2019note\endcsname

\bibitem[{Yoon \& Langer(2005)}]{yoonEvolutionRapidlyRotating2005}
Yoon, S.-C. \& Langer, N. 2005, A\&A, 443, 643 \csname
  yoonEvolutionRapidlyRotating2005link\endcsname~\csname
  yoonEvolutionRapidlyRotating2005note\endcsname

\bibitem[{Yoon {et~al.}(2006)Yoon, Langer, \&
  Norman}]{yoonSingleStarProgenitors2006}
Yoon, S.-C., Langer, N., \& Norman, C. 2006, A\&A, 460, 199 \csname
  yoonSingleStarProgenitors2006link\endcsname~\csname
  yoonSingleStarProgenitors2006note\endcsname

\end{thebibliography}
%---------------------------------------------------------------------
\begin{appendix}
\section{Energy transfer location} \label{app:et}
Because our ET formalism is implemented explicitly, and the radius of a stellar model can change during the solver iterations, which is also coupled to mass added or subtracted due to MT, a non negligible mismatch between the location of energy input and the RL can occur.
In particular, when mass is removed due to MT, the ET location is then evaluated on the remaining cells, without those cells being adjusted since the previous step.
Therefore, in phases with a high MT rate, it is possible that all mass above the RL is removed in one step (although after the solver performs the integration, the star will expand due to thermal relaxation).
This means that the location of ET in terms of radius is rather poorly defined.
To mitigate this, instead of choosing directly to put energy in cells with radii
\begin{equation}
    r \in [1.00, 1.01]R_{\rm RL},
\end{equation}
we opt to do following.
At the beginning of the evolutionary step, we evaluate:
\begin{equation}
    x_{\rm depth} = 1-m(r=R_{\rm RL})/M
\end{equation}
as the depth fraction of the current RL in mass coordinate.
Then during the next integration, the location of energy input is cells
\begin{equation}
    r \in [1.00, 1.01]r(x=x_{\rm depth}).
\end{equation}
This treatment mitigates mismatches in ET locations introduced by MT during evolutionary steps to the degree that the depth fraction changes from step to step.
\section{Convergence study}\label{app:convergence}
This appendix details the convergence study we used to validate our stellar model with and without energy transfer.
We have implemented the ET method in both MESA version r15140, as well as version r22.11.1.
\par
\subsection{Explicit evaluation}\label{ssec:expl}
In MESA r15140, both the tidal deformation corrections of \citetalias{fabryModelingOvercontactBinaries2022} and the ET of Sect. \ref{ssec:et} are evaluated explicitly in the differential equation solver, meaning only information of the previous evolutionary step is used.
\par
\begin{figure}
    \centering
    \includegraphics{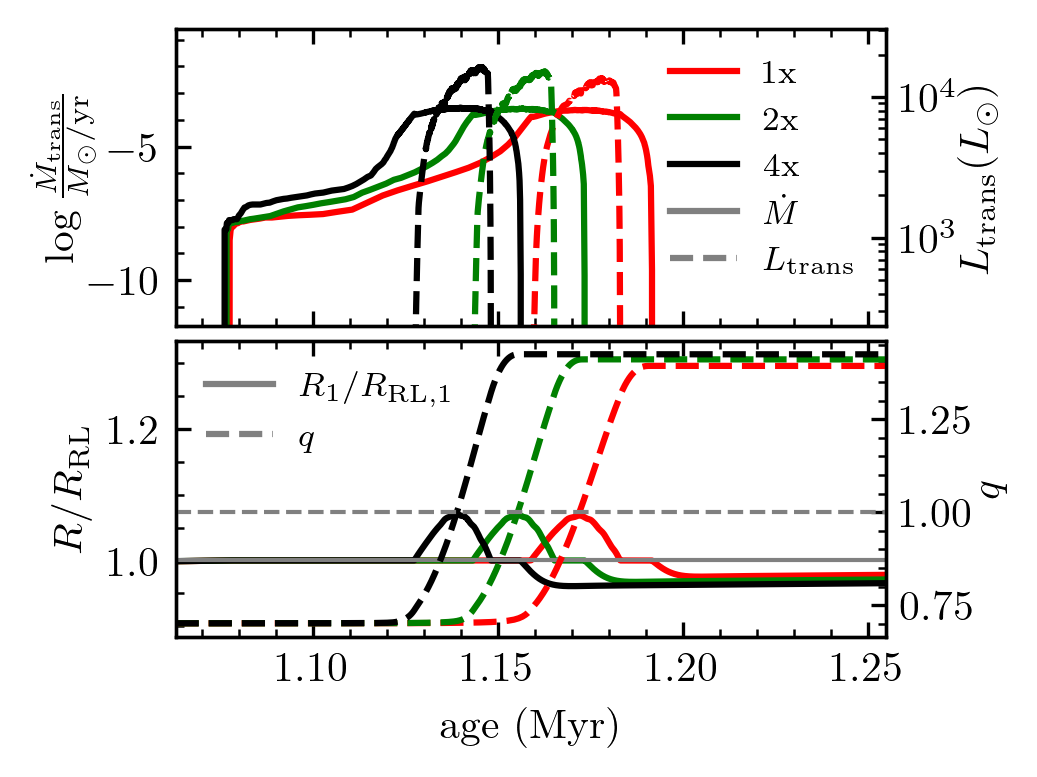}
    \caption{Evolution of the MT rate and ET rate (top panel) and the fractional RL radius and mass ratio (bottom panel) during fast case A MT for a $20 + 14 M_\odot$ binary, computed using the explicit evaluation of the tidal deformation corrections in MESA r15140. The color indicates what time resolution scale has been used.}
    \label{fig:convergence15140}
\end{figure}

We perform a time resolution study in which we take a base template, and increase the time resolution by successive factors of two.
Figure \ref{fig:convergence15140} shows the evolution of the MT rate, ET rate, the mass ratio and the donor radius during the fast case A MT phase.
While the time of the onset of MT, which coincides with the donor reaching the RL, remains roughly constant for the different resolutions, the MT rate evolves significantly differently.
For higher time resolutions, the MT rate is higher, resulting in slightly shorter contact phase and a higher final mass ratio.
This effect is independent of our implementation of ET since the changes occur before the contact phase (signified by the bump in the donor radius).
Furthermore, these models do not yet converge to a certain value for the quantities of interest, which are the contact phase duration and the mass ratio after the MT phase.
This means we would need to increase the time resolution further to get to a convergent solution.
This is prohibitive however keeping in mind the aim is to run a full population of models in later work.
We therefore elect to implement the tidal deformation corrections in MESA version 22.11.1 to allow for implicit evaluation during integrations.
\par

\subsection{Implicit evaluation}
In MESA r22.11.1, infrastructure is provided to supply numerically approximated derivatives necessary for the solver to implicitly evaluate the tidal deformation corrections during an integration step.
Generally, implicit evaluation is beneficial for numerical stability, and allows for longer time steps.
Note that the amount of energy to be transferred and its location in the stellar structure is still evaluated explicitly, as detailed in Sect. \ref{ssec:et}.
\par
\begin{figure}
    \centering
    \includegraphics{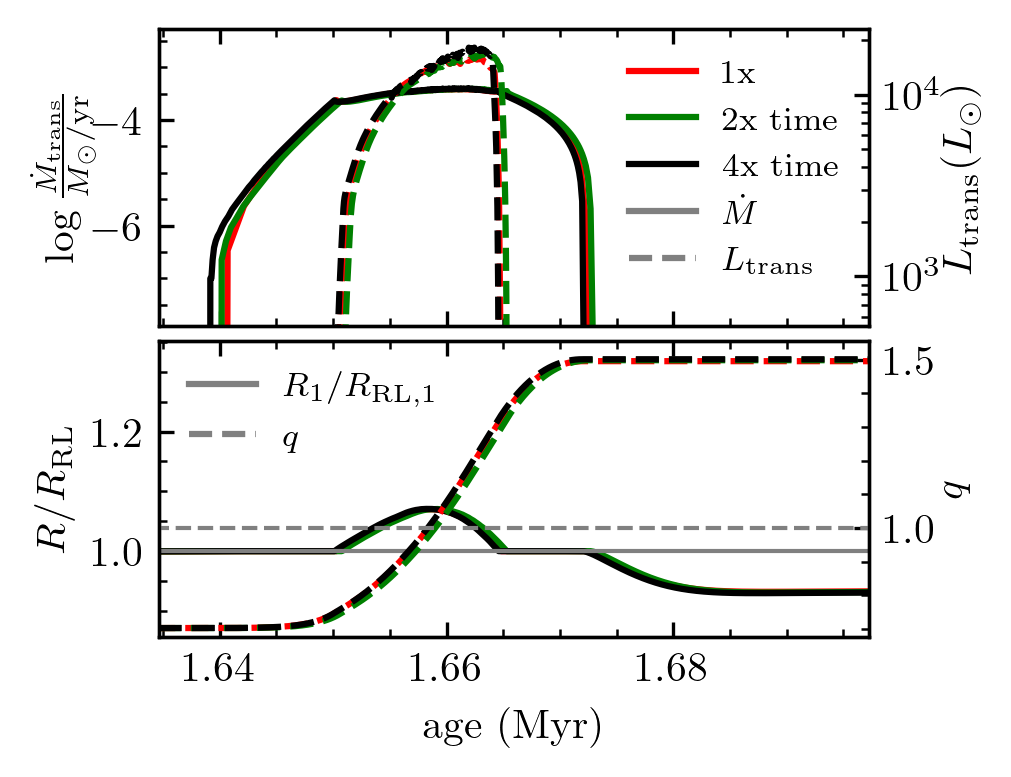}
    \caption{Same as Fig. \ref{fig:convergence15140}, but now the models were computed using implicit evaluation of the tidal deformation corrections in MESA r22.11.1.}
    \label{fig:convergence22}
\end{figure}

We do a similar resolution study as in Sect. \ref{ssec:expl} of a $20+14M_\odot$ binary, were we progressively increase the time resolution between simulations.
Compared to version r15140, we observe much better convergence of the MT rate evolution, as pictured in Fig. \ref{fig:convergence22}.
\par
\begin{figure}
    \centering
    \includegraphics{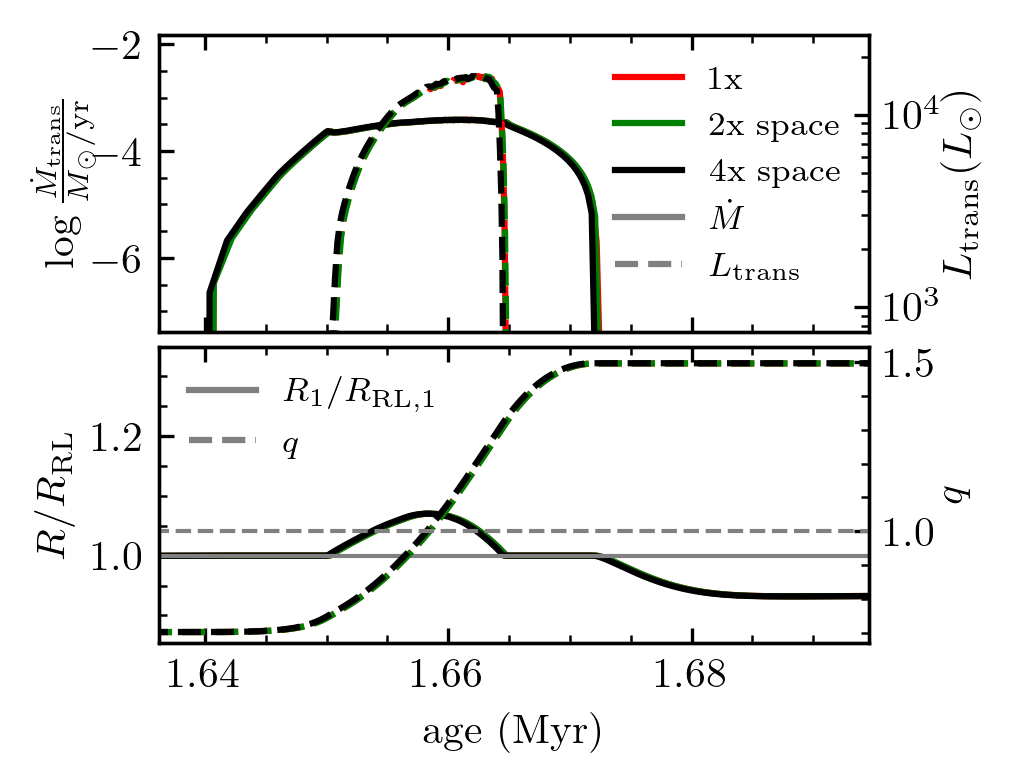}
    \caption{Same as Fig. \ref{fig:convergence22}, but now the color indicates the increase in spacial resolution.}
    \label{fig:spaceconvergence}
\end{figure}

Finally, we perform a spacial resolution convergence study for the same initial conditions as before.
Figure \ref{fig:spaceconvergence} shows that our models are well converged in the spacial coordinate already at the 1x level.
Even during the contact phase where ET occurs, and a sharp increase in luminosity is introduced, the models depend very weakly on the spacial resolution.
\par
\end{appendix}
\end{document}